\RequirePackage{amsmath} 
\documentclass[]{fairmeta}

\usepackage[utf8]{inputenc} 
\usepackage[T1]{fontenc}    
\usepackage{hyperref}       
\usepackage{url}            
\usepackage{booktabs}       
\usepackage{amsfonts}       
\usepackage{nicefrac}       
\usepackage{microtype}      
\usepackage{xcolor}         

\usepackage{pgfplots}
\usepackage{tikz}
\usetikzlibrary{arrows,automata,positioning}
\usepackage[inline]{enumitem}

\usepackage{pifont}
\newcommand{\cmark}{\ding{51}}%
\newcommand{\xmark}{\ding{55}}%

\definecolor{lightblue}{rgb}{0.88, 0.96, 1}
\definecolor{av_comment}{RGB}{255, 128, 0}

\usepackage[normalem]{ulem}
\usepackage{colortbl}

\newcommand{\ab}{Audiobox}
\newcommand{\abssl}{Audiobox SSL}
\newcommand{\absd}{Audiobox Sound}
\newcommand{\absp}{Audiobox Speech}

\newcommand{\datamixall}{Mix-185K}
\newcommand{\dataspbook}{SP-book-60K}
\newcommand{\dataspmulti}{SP-multi-100K}
\newcommand{\datasdtag}{SD-tag-6K}
\newcommand{\datasdcap}{SD-cap-150}

\newcommand{\boldparagraph}[1]{\noindent{\bf #1:}}

\newcommand{\ci}[1]{{\tiny $\pm$ #1}}

\newif\ifdraft
\drafttrue 

\ifdraft
    
    \newcommand{\whr}[1]{\textcolor{magenta}{\sout{#1}}}
    \newcommand{\whc}[1]{\textcolor{magenta}{[WN: #1]}}
    \newcommand{\mlc}[1]{\textcolor{magenta}{[ML: #1}}
    \newcommand{\avc}[1]{\textcolor{av_comment}{[AV: #1]}}
    \newcommand{\ava}[1]{\textcolor{av_comment}{#1}}

    \newcommand{\avt}[1]{\textcolor{blue}{[AV: #1]}}
    \newcommand{\atj}[1]{{\color{violet}[AT: #1]}}
    
    \newcommand{\bsc}[1]{\textcolor{blue}{[BS: #1]}}

    \newcommand{\hleditc}[1]{\textcolor{red}{#1}}

\else
    
    \newcommand{\whr}[1]{}
    \newcommand{\whc}[1]{}
    \newcommand{\mlc}[1]{}
    \newcommand{\avc}[1]{}
    \newcommand{\ava}[1]{#1}
    \newcommand{\avt}[1]{}
    \newcommand{\atj}[1]{}
    \newcommand{\bsc}[1]{}

    \newcommand{\hleditc}[1]{}

\fi

\title{\ab{}: Unified Audio Generation\\with Natural Language Prompts}

\author[*]{Apoorv Vyas}
\author[*]{Bowen Shi}
\author[*]{Matthew Le}
\author[*]{Andros Tjandra}
\author[*]{Yi-Chiao Wu}

\author[]{Baishan Guo}
\author[]{Jiemin Zhang}
\author[]{Xinyue Zhang}
\author[]{Robert Adkins}
\author[]{William Ngan}
\author[]{Jeff Wang}

\author[]{Ivan Cruz}
\author[]{Bapi Akula}
\author[]{Akinniyi Akinyemi}
\author[]{Brian Ellis}
\author[]{Rashel Moritz}
\author[]{Yael Yungster}
\author[]{Alice Rakotoarison}
\author[]{Liang Tan}
\author[]{Chris Summers}
\author[]{Carleigh Wood}
\author[]{Joshua Lane}
\author[\dagger]{Mary Williamson}
\author[\dagger]{Wei-Ning Hsu}

\affiliation[]{Audiobox Team, Fundamental AI Research (FAIR) at Meta}

\contribution[*]{Research team, equal contribution}
\contribution[\dagger]{Research and engineering leadership, equal contribution}

\abstract{Audio is an essential part of our life, but creating it often requires expertise and is time-consuming. Research communities have made great progress over the past year advancing the performance of large scale audio generative models for a single modality (speech, sound, or music) through adopting more powerful generative models and scaling data. However, these models lack controllability in several aspects: speech generation models cannot synthesize novel styles based on text description and are limited on domain coverage such as outdoor environments; sound generation models only provide coarse-grained control based on descriptions like ``a person speaking'' and would only generate mumbling human voices.
This paper presents \textsc{\ab{}}, a unified model based on flow-matching that is capable of generating various audio modalities. We design description-based and example-based prompting to enhance controllability and unify speech and sound generation paradigms. We allow transcript, vocal, and other audio styles to be controlled independently when generating speech. To improve model generalization with limited labels, we adapt a self-supervised infilling objective to pre-train on large quantities of unlabeled audio. 
\textsc{\ab{}} sets new benchmarks on speech and sound generation (0.745 similarity on Librispeech for zero-shot TTS; 0.77 FAD on AudioCaps for text-to-sound) and unlocks new methods for generating audio with novel vocal and acoustic styles. 
We further integrate Bespoke Solvers, which speeds up generation by over 25 times compared to the default ODE solver for flow-matching, without loss of performance on several tasks.}
\correspondence{Apoorv Vyas \email{vyasapoorv@meta.com}, Wei-Ning Hsu  \email{wnhsu@meta.com}}

\metadata[Demo]{\url{https://audiobox.metademolab.com/}}

\begin{document}
\maketitle

\begin{figure}
    \centering
    \includegraphics[width=0.8\linewidth]{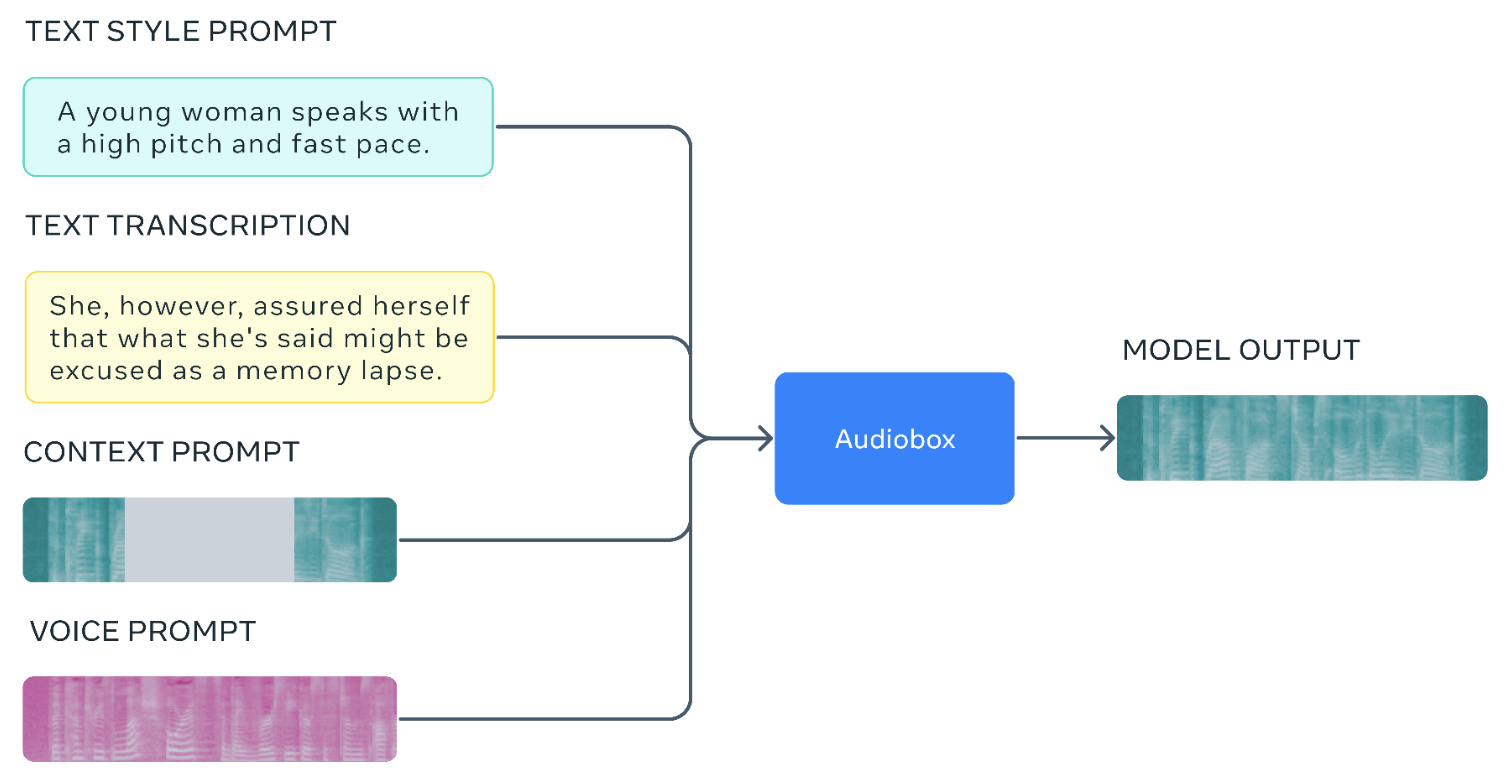}
    \caption{Audiobox model diagram}
    \label{fig:audiobox}
\end{figure}

\section{Introduction}

\boldparagraph{Why building audio generative models}
Audio is a key component in creating many forms of content, such as movies, podcasts, audiobooks, and Ads. However, audio creation is time-consuming and requires various expertise, such as voice acting, music composing and performing, Foley sound effect creation, and sound engineering. This imposes a great barrier to entry for the general public, making it hard for people to become audio creators. Even for professionals, performing these tasks can still take a lot of time and resources, limiting their productivity. Developing audio generative models that are generalizable, controllable, and high quality can bring transformative changes to the audio creation process, improving the efficiency of the professionals as well as unleashing the creativity for everyone.

\boldparagraph{Progress of audio generative models}
Recently, researchers have made significant progress advancing audio generative models. Speech generative models can mimic any vocal style using audio prompts that are as short as three seconds~\citep{Wang2023NeuralCL, shen2023naturalspeech, le2023voicebox, spear-tts}, infill a portion of speech to remove transient noise or edit words for any speaker~\citep{le2023voicebox, shen2023naturalspeech}, synthesize foreign languages in anyone's voice~\citep{zhang2023speak, le2023voicebox}, and create dialogues~\citep{borsos2023soundstorm}. 
Music generative models can create music in various styles using a short text description~\citep{schneider2023mo,huang2023noise2music,agostinelli2023musiclm,copet2023simple} and infill a portion of music~\citep{li2023jen}.
Sound effect generative models follows a similar paradigm. They are capable of creating and infilling complex acoustic scenes like \textit{``birds chirping and water dripping with some banging in the background''} given a text description~\citep{yang2023diffsound,kreuk2022audiogen,huang2023make,ghosal2023text,liu2023audioldm,liu2023audioldm2}. Recent models also extends to more general editing, such as removal or addition of sound events with natural language instructions~\citep{wang2023audit,liu2023separate}.

\boldparagraph{Limitation of existing models} Existing audio generative models are still limited in controllability and generalizability. First, the real world audio content often contain a mix of speech, music, and sound effects. However, existing audio generative models are mostly modality-specific, which only generate either speech, music, or sound effects. In particular, existing large scale speech generative models~\citep{Wang2023NeuralCL,le2023voicebox,shen2023naturalspeech} are trained mostly on audiobooks~\citep{zen2019libritts,Kahn2019LibriLightAB,Pratap2020MLSAL}, which lacks diversity compared to truly in-the-wild data such as AudioSet~\citep{gemmeke2017audio} in terms of expressivity (e.g., non-verbal sounds like coughing, screaming, laughing) and acoustic conditions (e.g., urban, rural, public indoor, stadiums). These models can only generate audio of limited styles and do not capture the correlation between different audio modalities.

On the other hand, there is a discrepancy between speech and sound/speech generation paradigm. Recent speech generation models mostly use example-based control, where an audio sample of the target style is provided and the style control is more precise; in contrast, description-based control is adopted for music and sound generation, where the model can create novel styles based on natural language prompts. Both approaches have their strengths and weaknesses, but such a discrepancy prevents development of unified models that enjoy the best of both worlds.

Last but not least, existing sound generation models only provide coarse control such as \textit{``a man is speaking''} when generating speech. Existing datasets do not offer finer-grained captions that characterizes vocal styles in greater details, such as 
\textit{``A middle aged woman from the American South is speaking over the phone in a passionate voice. She speaks in at a fast pace with a high pitch.''} Neither do these models enable transcript input to controlling the textual content. Hence, these models can only generate mumbling speech.

Due to a lack of consideration in the language-guided generation of speech within a natural setting, designing proper objective evaluation metrics for such universal models remains an open question that has not been fully addressed by prior works. In objective evaluation, previous speech-oriented studies~\cite{guo2023prompttts,leng2023prompttts,yang2023instructtts} often adopt ad-hoc evaluation metrics (e.g., accuracy of pre-defined attributes), making it challenging to generalize to free-form instructions. The joint audio-text embedding network (e.g., CLAP~\cite{wu2023large}), widely utilized in text-to-audio generation, is tailored to sound events and frequently falls short in capturing intricate attributes such as accents in speech (see \cref{sec:ab_spcap}).

\boldparagraph{Goals and overview of our model}
To tackle these problems, there are three key objectives of this work. 
First, we aim to build a unified model for sound and speech in order to generate a wider variety of real-world audio, which is often a mix of both.
Second, we want to improve controllability for creating novel styles through enabling multiple input methods, using either reference audio, text description, or a combination of both.
Last but not least, to improve model generalization, we want to scale training data and utilize data with different level of supervision.

To that end, we present the \ab{} framework. Audiobox is built upon Voicebox~\citep{le2023voicebox} and SpeechFlow~\citep{liu2023generative}, which are flow-matching based models for transcript-guided speech generation and self-supervised speech pre-training, respectively.
To facilitate data scaling and development of downstream models, we first adopt the SpeechFlow pre-training method and pre-train a unified model using large quantities of unlabeled speech, music, and sound effects, referred to as \textsc{\abssl{}} (\cref{sec:ab_ssl}). 
To validate the effectiveness of the unified pre-trained model, we fine-tune \textsc{\abssl{}} for transcript-guided speech generation (\textsc{\absp{}}, \cref{sec:ab_sp}) and description-guided sound generation (\textsc{\absd{}}, \cref{sec:ab_sd}), showing significant improvements from prior studies.

Combining the best of both worlds, we present \textsc{\ab{}}, the unified model for sound and speech generation in \cref{sec:ab_uni}. It bridges the gap between sound and speech generation by enabling natural language prompts for holistic style control, and furthers disentangled speech control with voice prompts. 
Our joint model achieves unprecedented controllability for universal audio generation and superior versatility with additional capabilities on top of what Voicebox offers. \textsc{\ab{}} outperforms existing domain specific models on multiple tasks and is close to \textsc{\absp{}} and \textsc{\absd{}} on their corresponding benchmark tasks.

To facilitate the evaluation of \ab{} and advance research in text-guided universal audio generative models, we propose Joint-CLAP, trained on both sound and speech description data. In comparison to CLAP~\cite{wu2023large}, Joint-CLAP significantly outperforms CLAP in retrieving description-based speech, and the text-to-audio similarity exhibits a stronger correlation with human judgment.

Orthogonally, to improve performance-efficiency trade-off, we integrate Bespoke Solver, a novel post-training inference optimization methods for flow-matching models. With Bespoke Solver, our models are able speed up by 25x compared to using the adaptive step size dopri5 solver without loss of performance.

As generative models become more powerful and essential parts of everyone's life, it is more important than ever to conduct research responsibly and mitigate potential risks. We conducted a series of study demonstrating the fairness is achieved through better representing voices of different demographic groups with data scaling. We also validate the effectiveness of a recent watermarking system~\citep{seamlessv2}, showing the verification is highly effective and robust to adversarial perturbation. 

\section{Related Work}
This paper is related to a large body of work on large scale generative modeling for audio. As the focus of this work is on universality and controllability, we first discuss controllable generation for modality specific models and then compare with recent studies on universal models that can perform multiple tasks or generate audio in multiple modalities and domains. For the rest of the paper, we will refer to speech, sound, music as different \textit{audio modalities}, and within modality style variation, such as read speech, spontaneous speech, conversational speech, as different \textit{domains}.

\boldparagraph{Large scale in-context text-to-speech generative models}
Over the past few months, there has been significant progress in developing large scale speech generative models~\citep{Wang2023NeuralCL, shen2023naturalspeech, spear-tts, le2023voicebox, yang2023uniaudio, borsos2023soundstorm} that are trained on in-the-wild data at the scale of close to 100K hours~\citep{Kahn2019LibriLightAB,Pratap2020MLSAL} with minimal supervision, which leads to much better generalization for synthesizing unseen speech styles in a zero-shot fashion.
These models are in sharp contrast to conventional regression-based models such as~\citet{Ren2020FastSpeech2F, Shen2017NaturalTS, lancucki2021fastpitch}, which are trained on highly curated datasets~\citep{Yamagishi2019CSTRVC} containing clean audio, limited style variation, and extensive labels (e.g., speaker and emotion labels).

The key to successful data scaling in recent work is the adoption of powerful generative models that can capture highly stochastic input-output relationships.
For example, VALL-E~\citep{Wang2023NeuralCL} adopt the token-based autoregressive language modeling approach, which converts speech into discrete tokens with a neural codec model~\citep{Defossez2022HighFN} and formulate text-to-speech (TTS) as a conditional language modeling problem given a transcript and an audio prompt (the first few seconds of the target speech).
NaturalSpeech2~\citep{shen2023naturalspeech} and Voicebox~\citep{le2023voicebox} adopt non-autoregressive diffusion~\citep{Ho2020DenoisingDP} and conditional flow-matching models~\citep{flow-matching}. Given a transcript and an audio context (the audio surrounding the target speech), these models iteratively transform a noise sampled from a simple prior to speech, represented as learned latent features or mel spectrograms.

At the high level, VALL-E performs transcript-guided speech continuation while NaturalSpeech2 and Voicebox perform transcript-guided speech infilling. 
These models are trained with only transcript supervision, which facilitates data scaling. The \textit{style} of the generated audio is controlled through the audio prompt or audio context. 
Note that the style refers to not only voice, but everything other than transcript, including prosody, emotion, acoustic environment, channel, noise, etc. This can be understood as a form of \textit{in-context learning}: 
because the audio style tends to be coherent within an utterance, these models learn to infer the style of the target based on its context. 
In turn, it enables generalization to unseen style, such that speech of any style can be generated by conditioning on an audio prompt/context of the desired style.

While the in-context style transfer paradigm is powerful, it also possesses several limitations in terms of controllability. 
First, audio prompt is the only input mechanism of controlling the audio style. Users cannot provide a descriptive text, such as ``a young man speaking with a happy tone in an auditorium'' to create diverse speech matching the description, whereas this feature is commonly supported and widely enjoyed for image~\citep{ramesh2022hierarchical,rombach2022high}, music~\citep{agostinelli2023musiclm}, and sound~\citep{kreuk2022audiogen} generation. 
Second, disentangled style control is not enabled with the paradigm, where voice and other attributes, such as emotion and acoustic condition, can be controlled independently. This feature is often desired as exemplified in earlier work where emotion and voice can be controlled independently~\citep{Hsu2018HierarchicalGM, kulkarni2021improving, nguyen2023expresso}.

\boldparagraph{Natural language style prompting for controllable speech generation}
Studies on controllable speech generation aims to develop models which can generate speech of many different domains and provide input methods for disentangled, flexible, and accurate control.
Earlier models often enable control over only a small number of attributes (e.g., speaker and emotion) with a fixed number of options (e.g., happy/sad/neutral for emotion) through one-hot vectors~\citep{nguyen2023expresso}. Such methods are difficult to generalize as it is difficult to represent many speech attributes, such as audio quality, acoustic environment, with one-hot vectors. Nor could information such as ``a speaker starts with a slow pace and speeds up'' be accurately represented.
In-context TTS~\citep{Wang2023NeuralCL} models greatly improves domain coverage, but has the limitation on flexibility and disentangled control described above.

To address the limitation, several recent studies also propose to control speech style through natural language prompts. 
InstructTTS~\citep{yang2023instructtts} and PromptTTS~\citep{guo2023prompttts} are the two earliest works. 
They are trained on small scale data with mainly emotion variation and limited number of speakers (7 for InstructTTS and 2 for PromptTTS synthetic setup). 
In particular, InstructTTS collects human descriptions for 44 hours of speech focusing on only the emotion and a separate speaker ID input is used as model input. Therefore, the natural language prompt is only used for controlling the emotion.
PromptTTS recruits human annotators to write descriptions to given four to five attribute labels (emotion, gender, volume, speed, and pitch; emotion label is not available for the real data), and trains models on 2-voice synthetic data as well as LibriTTS~\citep{zen2019libritts}.
Because the descriptions of PromptTTS are created based on attribute labels instead of speech samples, these descriptions do not contain additional information compared to the labels and theoretically does not enable finer grained attribute control.

PromptTTS2~\citep{leng2023prompttts} is a concurrent work which improves upon PromptTTS in two aspects. First, it proposes a automatic description creation pipeline based on speech attribute labeler and large language models, which enables scaling to training on 44K hours of audiobook data. Second, PromptTTS2 adopts a diffusion model to capture the one-to-many relationship given input (transcript and description), whereas PromptTTS adopts a regression model assuming deterministic mapping. Nevertheless, similar to PromptTTS, all the descriptions PromptTTS2 create are derived from four categorical attributes with two to three options each (total 54 combinations). Hence, PromptTTS2 does not provide finer grained control than PromptTTS and has limited coverage on the attributes it can control via natural language prompt.

\boldparagraph{Large scale general-domain models for sound and music generation}
Text-to-sound~\citep{kreuk2022audiogen} and text-to-music~\citep{schneider2023mo} are the emerging paradigms for general-domain sound and music generation, in contrast to earlier studies that generate finite sound effects~\citep{donahue2018adversarial} or instruments~\citep{huang2018music}. The text here refers to a holistic description of the target audio, such as \textit{``A child shouts while an emergency vehicle siren sounds with the horn blowing.''}~\citep{kim2019audiocaps} and \textit{``The low quality recording features a ballad song that contains sustained strings... It sounds sad and soulful, like something you would hear at Sunday services.''} for music~\citep{agostinelli2023musiclm}. 

Similar to speech generation, the recent progress can be largely attributed to the advancement in generative models for continuous data~\citep{Ho2020DenoisingDP, huang2023noise2music, liu2023audioldm} and audio tokenizers~\citep{Zeghidour2022SoundStreamAE, Defossez2022HighFN, kreuk2022audiogen, copet2023simple, agostinelli2023musiclm}, which enables modeling methods capable of capturing highly stochastic conditional distributions of audio given descriptions for general domain sound/music data. 

A key limitation of these models is the ability to control transcript and generate intelligible speech or vocals. These models only take a description as input, which does not specify the transcript when speech is presented. Hence, generating samples with prompts like ``a person speaking'' often results in speech-like mumbling sound with unintelligible content~\citep{liu2023audioldm}. In other words, these models does not offer an input for users to control transcript, and have not learned language models that allow it to construct and synthesize meaningful sentences given only the description.

\boldparagraph{Unified model for audio generation}
With the great progress made in developing general-domain models for each audio modality, researchers also start exploring unified model that can generate audio beyond a single modality and perform multiple generative tasks. Such a model could potentially learn from different sources of supervision and benefit from knowledge transfer across tasks. There are three concurrent studies that are related to this work.

UniAudio~\citep{yang2023uniaudio} focuses on building a single model that can perform multiple tasks, including text-to-music, text-to-sound, and in-context TTS and natural language style prompted TTS. 
It follows the VALL-E~\citep{Wang2023NeuralCL} framework, which tokenizes audio and serializes conditioning input and output audio tokens for training a conditional token-based language model. 
It is trained on the same speech descriptions collected by PromptTTS, which inherits the same limitations in terms what attributes and how granular they can be controlled through natural language prompts as discussed earlier. 

VoiceLDM~\citep{lee2023voiceldm} is the most related work. It introduces a transcript input to AudioLDM~\citep{liu2023audioldm} and controls style through text description embedded with a frozen Contrastive Language-Audio Pre-training (CLAP) model~\citep{wu2023large}. During training, CLAP embedding from audio is used for conditioning. VoiceLDM is trained on datasets with rich acoustic variation, and hence is capable of generating speech in diverse acoustic environments. However, the performance in terms of controllability is bounded by the pre-trained CLAP model. Since the CLAP model are trained on audio-caption pairs focus on sound events, the embedding only encodes very coarse information regarding speech attributes. Furthermore, VoiceLDM also follows the sound generation paradigm which always generate audio clips of a fixed size (10 seconds), which is not ideal for speech generation that have variable length in general. Finally, despite that the model can generate non-speech sounds when conditioned on empty transcripts, the performance of sound generation lags behind state-of-the-art models by a large margin.

AudioLDM 2~\citep{liu2023audioldm2} presents a two-stage model that is applicable to speech, sound, and music generation. It is comprised of a deterministic auto-regressive model that maps conditioning input (e.g., CLAP-embedded audio, description, transcript, image) to semantic features sequence, and a diffusion model which mapping semantic to acoustic features. 
The structure is similar to SPEAR-TTS~\citep{spear-tts} but with different modeling methods and representations for each stage. Hence, similarly it can leverage unlabeled audio for training the second stage model. 
While AudioLDM 2 presents a unified framework, empirically separate models for speech and sound/music generation are trained, as the authors noted that different model architecture hyperparameters are required for different modalities. 
\section{Background}

This work is heavily built upon the training objective and model architecture of Voicebox~\citep{le2023voicebox}, and the self-supervised objective of SpeechFlow~\citep{liu2023generative}. Both studies adopt conditional flow-matching~\citep{flow-matching} as the modeling backbone, which is a powerful non-autoregressive generative model for continuous data. We provide a technical overview here.

\boldparagraph{Conditional flow-matching}
Conditional flow-matching (FM)~\citep{flow-matching} is a novel generative modeling method derived from the continuous normalizing flow~\citep{cnf} framework. It models the paths that transform samples from a simple prior distribution $p_0$ to the corresponding samples from the complex data distribution $p_1$ in a continuous manner. We use \textit{flow step} $t$ to describe the progress of transformation, where the prior is at $t=0$ and the data is at $t=1$. 

The training objective of FM resembles the objective diffusion models~\citep{Ho2020DenoisingDP}: during training, given a sample $x_1$ drawn from the data distribution, a random flow step $t\sim\mathcal{U}[0, 1]$ is sampled, and a noisy version of the data $x_t$ as well as its derivative $v_t=dx_t/dt$ for the chosen condition path are computed. A FM model $u$ is trained to predict the derivative $v_t$ given $t$ and $x_t$. 
During inference, to draw a sample $x_1$ from the learned data distribution, a sample $x_0$ is first drawn from the prior distribution, and then the ordinary differential equation (ODE) solver is used to estimate $x_1$ given $x_0$ and the derivative parameterized by the FM model through integration. Trade-off between accuracy of $x_1$ estimation and speed can be flexibly selected by configuring the ODE solver.

At a high level, FM subsumes diffusion models, which correspond to specific paths of the transformation. The authors of \citet{flow-matching} presented an alternative called optimal transport (OT), which are conditional paths with constant directions and speeds. It is arguably easier to learn and can be more accurately estimated by the ODE solver with fewer steps. The OT path results in better training and inference efficiency as empirically verified in \citet{flow-matching} and \citet{le2023voicebox}. 

Given a sample $x_1$ and a flow-step $t$, with the OT conditional path we have $x_t = (1-(1-\sigma_{min})t)x_0 + tx_1$ and $v_t = x_1 - (1 - \sigma_{min}) x_0$, where $x_0$ is drawn from the prior distribution $N(0, I)$ and $\sigma_{min}$ is a small value ($10^{-5}$). The FM model $u$ minimizes:
\begin{equation}
    \mathbb{E}_{t, x_1, x_0} || u(x_t, t) - v_t ||^2.
\end{equation}

\boldparagraph{Voicebox}
Voicebox~\citep{le2023voicebox} is a conditional generative model based on FM which additionally conditions on frame-aligned phonetic transcript and masked audio for audio prediction, and conditions on phonetic transcript and masked duration sequence for phone duration prediction. 
Audio is represented as 80-dimensional Mel spectrograms and are converted to waveform using a HiFi-GAN vocoder~\citep{kong2020hifi}. Duration sequence denotes the number of frames for each phoneme in the transcript. 

Voicebox adopts the Transformer~\citep{Vaswani2017AttentionIA} model with U-Net~\citep{ronneberger2015u} connections. Masked spectrogram (or masked duration), frame-aligned phone embeddings (or phone embeddings), and noisy audio $x_t$ (or noisy duration) are concatenated along the channel dimension and projected to the Transformer feature dimension. The flow step sinusoidal embedding is then concatenated with the project features along the time dimension, passed as input to the Transformer model. The Transformer output is then projected to 80 dimensions (or 1 dimension for duration) and predicts the derivative $v_t$.

It is a supervised model trained on 60K hours of audiobooks and achieves state-of-the-art performance on in-context text-to-speech synthesis that can mimic the audio style given a three second audio prompt. It is also high versatile due to the generality of transcript-guided infilling, where the model can perform transient noise removal, diverse style generation, speech editing, cross-lingual style transfer by simply forming transcript and audio inputs differently.

\boldparagraph{SpeechFlow}
SpeechFlow~\citep{liu2023generative} is a self-supervised framework based on FM with learns to infill speech given the audio context. This is equivalent to Voicebox without conditioning on transcripts. The self-supervised objective tackles label scarcity issues and enables the model to learn from large quantities of unlabeled speech the distribution of speech as well as the correlation between temporal segments within an utterance. 

Fine-tuning SpeechFlow with the same transcript-guided infilling objective as Voicebox shows superior performance and sample efficiency, matching style similarity of VALL-E~\citep{Wang2023NeuralCL} with only 10 hours of labeled data.
The pre-trained model also demonstrates promising improvements on other speech generation tasks, including source separation and speech enhancement.
It also enables parameter efficient fine-tuning like LoRA~\citep{hu2021lora} and fine-tuning with a much lower batch size, demonstrating the efficiency and reusability of self-supervised pre-train models.
\section{\textsc{\abssl{}}: Self-supervised Generative Audio Pre-training}\label{sec:ab_ssl}
Our first step is to develop \textsc{\abssl{}}, a foundation model that can be fine-tuned for any downstream audio generation tasks. Because labeled data are not always available or of high quality, and data scaling is the key to generalization, our strategy is to train this foundation model using audio without any supervision, such as transcripts, captions, or attribute labels, which can be found in larger quantities.

\subsection{Method}
We adapt \textsc{\abssl{}} from SpeechFlow, which was originally designed for generative speech pre-training. The same learning objective is also meaningful for general audio: through learning to infill, the model can also capture the temporal relationship of audio events (e.g., clock ticking sound at fixed time interval, approaching train producing sounds with increasing volume), and learns the distribution of general audio. Therefore, during supervised fine-tuning, a model does not need to learn what a natural audio sample sounds like, but only needs to learn aligning the label with the corresponding mode of distribution.

The original SpeechFlow model is trained to predict spectrograms and uses a HiFi-GAN model to generate waveform given spectrogram. However, HiFi-GAN does not generalize well to non-speech audio such as sound or music~\citep{lee2022bigvgan}. To tackle that, we train the model to predict latent features learned by an autoencoder. In particular, we use the dense Encodec~\citep{Defossez2022HighFN} features which are extracted prior to the residual quantization layer, which demonstrates good resynthesis quality in various audio modalities and  has been adopted for sound and music generation~\citep{kreuk2022audiogen, copet2023simple}. This is similar to the latent diffusion framework~\citep{rombach2022high} that is also adopted in NaturalSpeech2~\citep{shen2023naturalspeech}.

During training, the model is conditioned on fully masked features with probability $p_{\text{cond}}$.
With probability $1-p_{\text{cond}}$, a subset ($n_{\text{mask}})$ of frames are masked with minimum span length $l_{\text{mask}}$. The FM loss is computed only on masked frames. When a frame is masked, its features are set to $0$.

\subsection{Experimental Setup}

\boldparagraph{Training data}
We collect an large scale audio dataset that greatly increases the domain coverage, modality coverage, and quantities compared to previous large scale audio generative model studies~\citep{yang2023uniaudio, borsos2023soundstorm, Wang2023NeuralCL, liu2023audioldm2}, which leverage datasets ranging between 10K to 100K hours containing mostly speech from a single domain (e.g., audiobooks).

Specifically, our dataset includes over 160K hours of speech (primarily English), 20K hours of music and 6K hours of sound samples. The speech portion covers audiobooks, podcasts, read sentences, talks, conversations, and in-the-wild recordings including various acoustic conditions and non-verbal voices. To ensure fairness and a good representation for people from various groups, it includes speakers from over 150 countries speaking over 200 different primary languages. We refer to this set as ``\datamixall{}.''

\boldparagraph{Model and training}
We train a 24 layer Transformer \cite{Vaswani2017AttentionIA} with convolutional position embeddings \cite{baevski2020wav2vec} and symmetric bi-directional ALiBi self-attention bias \cite{Press2021TrainST}.  The model has 16 attention heads, 1024/4096 embedding/feed-forward network (FFN) dimension, and 330M parameters. We add UNet-style skip connections, where states are concatenated channel-wise and then combined using a linear layer.

The model is trained for 1 million updates with an effective batch size of 480K frames. For efficiency, samples are randomly chunked if they exceed 1,600 frames. We set $p_{\text{cond}}=0.1$, $n_{\text{mask}}\sim\mathcal{U}[70\%, 100\%]$, and $l_{\text{mask}}=10$. We use the Adam \cite{Kingma2014AdamAM} optimizer with learning rate 1e-4, linearly warmed up for 5k steps and linearly decayed over the rest of training.  For stability, we use gradient norm clipping with a norm threshold of 0.2.
\section{\textsc{\absp{}}: Scaling In-context Text-to-speech Synthesis}\label{sec:ab_sp}
In this section, we study the effectiveness of pre-training and fine-tuning data scaling for speech generation. We present \textsc{\absp{}}, which fine-tunes \textsc{\abssl{}} with the same transcript-guided speech infilling objective as Voicebox using transcribed speech. The resulting model can be applied to multiple downstream tasks just like Voicebox. 

\subsection{Method}

To incorporate the frame-aligned transcript $z$, we follow \cite{liu2023generative}.  Specifically, given the noisy Encodec features $x_t$ at the flow-step $t$, masked Encodec features $x_{\text{ctx}}$, we first concatenate $x_t$ and $x_{\text{ctx}}$ channel-wise and apply a linear project to get $x_h$. We then apply another linear layer to the frame-aligned transcript embeddings $z_{\text{emb}}$, and add this to the hidden state $x_h$. The resulting features are concatenated with the flow step sinusoidal embedding along the time dimension and fed to the Transformer as input. The Transformer output is projected and predicts the derivative $v_t$.

There are two different approaches to fine-tuning the model. The first one is low-rank adaptation (LoRA) ~\cite{hu2021lora}, where we add LoRA adapters to the linear input projection of each self-attention layer. With this approach, only the transcript embedding, projection parameters, along with the LoRA adapter parameters are optimized. The second approach is full fine-tuning, where all parameters are optimized together. \citet{liu2023generative} showed that LoRA achieves better performance when fine-tuning SpeechFlow on 960 hours of speech, but we suspect that full fine-tuning may prevail when we scale fine-tuning data.

In addition, many prior studies~\citep{le2023voicebox, Wang2023NeuralCL} represent transcripts as phoneme sequences and using the off-the-shelf Montreal Forced Aligner~\citep{McAuliffe2017MontrealFA} for aligning the training data. Instead, we represent transcript with raw characters, including punctuation and with true cases, and utilize the \textsc{SeamlessM4T v2} multilingual char-to-unit forced aligner presented in \citet{seamlessv2} adapted from RAD-TTS~\citep{shih2021rad}. This aligner is trained on large quantities of multilingual data and can align raw text with speech. There are several benefits with the replacement. First, it circumvents the need of phonemizers and avoids error propagation due to incorrect phonemization. Second, raw text preserves more information than phonemized text, such as casing (e.g., all caps for emphasis) and punctuation. Third, the \textsc{SeamlessM4T v2} aligner is much more robust than MFA and can handle multilingual/code-switching text, which enables easier extension to multilingual TTS systems and is more suitable for aligning challenging speech such as conversational and noisy samples.

Following~\citet{le2023voicebox}, we train a flow-matching duration model only with labeled data. It was shown in \citet{le2023voicebox} that FM duration model has better diversity compared to regression duration models. However, it is less stable and sometimes produces unnatural prosody. To alleviate the issue, we propose to average over a small number of duration sequences for stabilization, which empirically shows better trade-off between diversity and quality. The averaging operation is reasonable as duration distributions are relatively unimodal. When averaging more samples, it approaches the mean, which is the estimation produced by regression models.

\subsection{Task and Evaluation} \label{sec:ab_sp_task}

We consider the in-context TTS (also known as zero-shot TTS) task. In-context TTS aims to synthesize speech that resembles the audio style of the given an audio example which may be unseen during training. The audio style refers to not only voice, but everything other than transcript, such as prosody and acoustic condition. To perform the task, input raw/frame-level transcript is the concatenation of the raw/frame-level transcript of the audio example and the target raw/frame-level transcript, while the masked audio/duration is the concatenation of the example audio/duration and a mask for the speech/duration to be generated. We first sample duration sequence for the target raw transcript to create frame-level target transcript using the duration model, and then sample audio with the audio model.

The performance is measured in terms of style similarity, content correctness, and quality. A proxy automatic metric for style similarity is the cosine similarity between the audio prompt and the generated audio in some embedding space that reflects the audio style. WavLM-TDCNN~\citep{Chen2021WavLMLS} is commonly used for embedding~\citep{Wang2023NeuralCL,spear-tts,le2023voicebox}. \citet{le2023voicebox} advocates for reporting both similarity with respect to raw audio (SIM-orig) and to audio resynthesized from the same vocoder (SIM-resyn) for comparability across studies (SIM-orig). Content correctness can be approximated with the word error rate (WER) from some speech recognition model; however, WER can result from both synthesis error and recognition error, and hence is less reliable when numbers are close or when the target style is more difficult to recognize (e.g., accented speech, conversational speech, noisy speech). In this paper we use Whisper \texttt{large-v2} instead of HuBERT-L~\citet{Hsu2021HuBERTSS} used in prior studies~\citep{Wang2023NeuralCL,le2023voicebox} because the latter is less robust and has higher WER on real data for non audiobook domains.
Subjective evaluations are often used for assessing style similarity and audio quality, measured by mean opinion scores (MOS).

\subsection{Experimental Setup}
\label{sec:ab_sp_exp}
\boldparagraph{Training data}
We train \textsc{\absp{}} on a transcribed English subset of the speech data used for pre-training. The subset contains 100K hours of speech covering similar domains as the full set, which we refer to as ``\dataspmulti{}.'' We create the transcribed subset with the following pre-processing methods:

For unsegmented multi-speaker conversational datasets information, we first segment our dataset using PyAnnote diarization toolkit~\citep{Plaquet23, Bredin23} to create single speaker speech segments. 
For untranscribed speech, we transcribe data using two speech recognition models, Whisper \cite{radford2022robust} \texttt{large-v2} and \texttt{medium.en}. For each audio with unknown language, we additional use the Whisper \texttt{large-v2} model for language identification (LID). We then remove the utterances where the probability being English is lower than 50\% or the the word error rate (WER) between the transcriptions from the two models is greater than 50\%.

To create a similar text distributions across multiple datasets, we apply inverse text normalization to create true-cased and punctuated transcript for any dataset with normalized transcript using Whisper-punctuation library.\footnote{\url{https://github.com/jumon/whisper-punctuator}} It performs the task through constrained search where the produced transcript needs to match the original transcript after normalization.

\boldparagraph{Model and training}
We adopt the full fine-tuning method and train the audio model for 200K steps with an effective batch size of 240K frames. Samples are randomly chunked if they exceed 1,600 frames. Character embeddings are 128 dimensions. For each batch, audio is entire masked with probability 0.3; otherwise a contiguous chunk is masked where the chunk size 70\% to 100\% of the frames. The same optimizer, learning rate, scheduler, and gradient clipping as \textsc{\abssl{}} are used.

The duration model has 8 heads, 768/2048 embedding/FFN dimensions, 10 layers, with 40 dimension character embeddings. It is trained for 600K updates with an effective batch size of 120K frames. For each batch, duration is entirely masked with probability 0.2 and otherwise a chunk of 10\% to 100\% of the sequence length is masked. The rest of the optimization parameters are the same as the audio model.

\boldparagraph{Evaluation data and configuration}
For in-context TTS, three second prompts are used following \citet{Wang2023NeuralCL}. Voicebox uses the last three seconds of the reference as the prompt, which often contains a considerable amount of trailing silence. We instead use the last three seconds after removing the trailing silences based on the forced alignment for all experiments in this paper. Duration is estimated by averaging over five samples and following \citep{le2023voicebox} predicted silence at both ends are trimmed to 0.1 second max.

The \texttt{torchdiffeq}~\citep{torchdiffeq} package is used. By default, we use the midpoint solver with a step size of 0.0625, which invokes the derivatives being evaluated 32 times. When using classifier free guidance the model does 2 forward passes per evaluation, leading to a total of 64 calls to the model.  A guidance weight for classifier-free guidance~\citep{ho2022classifier} of 0.7 is applied.

Models are evaluated on five datasets representing different domains. 
\begin{enumerate*}[label=(\arabic*)]
    \item Librispeech test-clean (LS)~\citep{Panayotov2015LibrispeechAA}: audiobook recordings that are scripted and relatively clean. Following \citet{Wang2023NeuralCL}, we keep only samples between 4 to 10 seconds for evaluation to compare with prior studies. 
    \item CommonVoice v13.0 English test set (CV)~\citep{Ardila2019CommonVA}: sentences read by volunteers worldwide. It covers broader accents and are noisier compared to Librispeech. 
    \item Switchboard (SWBD)~\citep{godfrey1992switchboard}: a conversational speech corpus. We evaluate on a subset of 611 samples from 8 speakers.
    \item Expresso~\citep{nguyen2023expresso} (Expr) is a multispeaker expressive speech dataset covering 7 different speaking styles, which we evaluate on a subset of 999 samples. 
    \item An internal expressive and accented dataset (Accent): read sentences with speakers covering a wider range of accents and 10 emotions. We create a subset of 500 samples for evaluation.
\end{enumerate*}

\subsection{Main Results}
We compare \textsc{\absp{}} with several state-of-the-art in-context speech generation models. Voicebox, VALL-E, NaturalSpeech 2  (NS2), and YourTTS are trained on 60K, 60K, 44K, 600 hours of audiobooks respectively. UniAudio is trained on about 100K hours of audio, where speech accounts for 81K hours and are mostly audiobooks. Results are shown in \cref{tab:inctx_tts_spk,tab:ab_inctx_tts_mos}. 

\textsc{\absp{}} achieves a new best on style similarity (0.745 vs. 0.710 from UniAudio) on the audiobook domain test set (LS). More importantly, \textsc{\absp{}} drastically improves Voicebox on all other domains, with similarity improvement ranging from 0.096 to 0.156. The results suggest that \textsc{\absp{}} generalizes much better thanks to scaling data to cover more domains.
The subjective evaluations presented in \cref{tab:ab_inctx_tts_mos} again confirms that \textsc{\absp{}} transfers styles significantly better than the baselines, and generate audio with better quality.

\begin{table}[ht]
    \centering
    \caption{In-context TTS style similarity and content correctness. We cite \citet{yang2023uniaudio} for the NS2 results which are not in the original paper\citep{shen2023naturalspeech}. WER with $^*$ are computed using HuBERT-L ASR that is not comparable with the other numbers.}
    \label{tab:inctx_tts_spk}
    \resizebox{\linewidth}{!}{
    \begin{tabular}{c|c|cccccc|cccccc}
        \toprule
        & Sim-r $\uparrow$ & \multicolumn{6}{c}{Sim-o $\uparrow$} & \multicolumn{6}{c}{Word error rate (\%) $\downarrow$} \\
                    & LS  & LS & CV & SWBD & Expr & Accent & Avg & LS & CV & SWBD & Expr & Accent & Avg \\
        \midrule
        VALL-E      & 0.580 & - &- &- &- &- &-                              & 5.9$^*$ &- &- &- &- &- \\
        NS2         & 0.620 & - &- &- &- &- &-                              & 2.3$^*$ &- &- &- &- &- \\
        UniAudio    & 0.710 & - &- &- &- &- &-                              & 2.0$^*$ &- &- &- &- &- \\
        YourTTS     & -     & 0.455 & 0.312 & 0.291 & 0.290 & 0.366 & 0.343 & 6.8 & 10.4 & 11.8 & 9.5 & 4.0 & 8.5 \\
        Voicebox    & 0.696 & 0.674 & 0.477 & 0.452 & 0.487 & 0.563 & 0.531 & \textbf{2.6} & 7.9  & 10.6 & 7.2 & 2.1 & 6.1 \\
        \midrule
        \textsc{\absp{}} & \textbf{0.745} & \textbf{0.734}	& \textbf{0.607}	& \textbf{0.608}	& \textbf{0.603}	& \textbf{0.659}	& \textbf{0.642} & 3.2 & \textbf{3.7}  &  \textbf{9.1} & \textbf{3.2} & \textbf{0.9} & \textbf{4.0} \\
        \bottomrule
    \end{tabular}
    }
\end{table}

\begin{table}[ht]
    \centering
    \caption{In-context TTS style similarity and quality subjective evaluation}
    \label{tab:ab_inctx_tts_mos}
    \begin{tabular}{c|ccccc}
        \toprule
        \multicolumn{6}{c}{Style similarity MOS $\uparrow$} \\
        \midrule
        & {LS} & {CV} & {SWBD} & {Expr} & {Accent}\\

        YourTTS & {1.67 \ci{ 0.09}} & {1.61 \ci{ 0.09}} & {1.55 \ci{ 0.08}} & {1.41 \ci{ 0.07}} & {1.46 \ci{ 0.07}} \\
        Voicebox & {2.85 \ci{ 0.12}} & {2.66 \ci{ 0.13}} & {2.89 \ci{ 0.13}} & {2.42 \ci{ 0.13}} & {2.51 \ci{ 0.11}} \\
        \textsc{\absp{}} & {\textbf{3.88 \ci{ 0.11}}} & {\textbf{3.77 \ci{ 0.11}}} & {\textbf{3.63 \ci{ 0.12}}} & {\textbf{3.85 \ci{ 0.11}}} & {\textbf{3.77 \ci{ 0.11}}} \\

        \midrule
        \multicolumn{6}{c}{Quality MOS  $\uparrow$} \\
        \midrule
        & {LS} & {CV} & {SWBD} & {Expr} & {Accent} \\

        YourTTS & {1.89 \ci{ 0.10}} & {2.19 \ci{ 0.12}} & {1.57 \ci{ 0.08}} & {1.74 \ci{ 0.09}} & {1.92 \ci{ 0.10}} \\
        Voicebox & {3.70 \ci{ 0.11}} & {3.06 \ci{ 0.12}} & {2.94 \ci{ 0.12}} & {2.76 \ci{ 0.12}} & {3.38 \ci{ 0.12}} \\
        \textsc{\absp{}}  & {\textbf{4.11 \ci{ 0.08}}} & {\textbf{4.00 \ci{ 0.09}}} & {\textbf{3.74 \ci{ 0.09}}} & {\textbf{4.00 \ci{ 0.09}}} & {\textbf{4.22 \ci{ 0.07}}} \\
        \bottomrule
    \end{tabular}
\end{table}

\subsection{Ablation Study}\label{sec:ab_sp_ablate}
We present ablation studies in \cref{tab:inctx_ablate}. To understand the effect of data scaling, we create a subset containing 60K hours of audiobook speech referred to as ``SP-book-60K'', which is a subset of the 100K hour multi-domain speech we have (SP-multi-100K). 

We first compare the top two rows, which differ in the pre-training data and are both fine-tuned with LoRA. Results suggest that while WER remains similar, scaling pre-training data greatly improves style similarity, especially on domains not covered in the fine-tuning data (CV, SWBD, Expr, Accent). 
On the other hand, scaling fine-tuning data from SP-book-60K to SP-multi-100K does not appear to improve much on similarity. This potentially results from the fact that pre-training data is a superset of fine-tuning data, and hence fine-tuning has little to learn on style transfer and focuses on aligning transcript with speech. 

Comparing the third and the fourth row, we see that by fine-tuning the whole model, style similarity improves slightly and WER improves greatly on most of the domains (23\% to 43\% relative WER reduction). The only exception is on SWBD, which are 8kHz narrowband recordings that are likely less represented in the fine-tuning data. Finally, we compare the last two rows and confirm that using audio prompts without silence leads to drastic improvements on similarity on datasets which tend to have long trailing silences (CV, Accent), while overall maintaining the WER. This is because the silence is not informative for inferring the target style.

\begin{table}[ht]
    \centering
    \caption{Ablation study for in-context TTS. PT and FT data denote the data used for pre-training and fine-tuning repsectively. FT method denotes whether LoRA or full fine-tuning (full) is adopted. ``has sil'' denote whether the conditioned audio prompt contains silence.}
    \label{tab:inctx_ablate}
    \resizebox{\linewidth}{!}{
    \begin{tabular}{cccc|ccccc}
        \toprule
        & & & & \multicolumn{5}{c}{Sim-o $\uparrow$}\\
        PT data & FT data & FT method & has sil & LS & CV & SWBD & Expr & Accent \\
        \midrule
        \dataspbook{}  & \dataspbook{}   & LoRA & Y & 0.708 & 0.461 & 0.530 & 0.552 & 0.529 \\
        \datamixall{}     & \dataspbook{}   & LoRA & Y & 0.718 & 0.505 & 0.592 & 0.571 & 0.584 \\
        \datamixall{}     & \dataspmulti{} & LoRA & Y & 0.714 & 0.502 & 0.583 & 0.559 & 0.590 \\
        \datamixall{}     & \dataspmulti{} & full & Y & 0.720 & 0.508 & 0.556 & 0.603 & 0.596 \\
        \datamixall{}     & \dataspmulti{} & full & N & 0.734 & 0.607 & 0.608 & 0.603 & 0.659 \\ 
        \midrule\midrule
        & & & & \multicolumn{5}{c}{WER (\%) $\downarrow$}\\
        PT data & FT data & FT method & has sil & LS & CV & SWBD & Expr & Accent \\
        \midrule
        \dataspbook{} & \dataspbook{} & LoRA & Y & 4.4 & 4.4 & 8.7 & 4.2 & 1.5 \\
        \datamixall{}     & \dataspbook{}   & LoRA & Y & 3.8 & 4.7 & 8.9 & 3.9 & 1.4 \\
        \datamixall{}     & \dataspmulti{} & LoRA & Y & 3.8 &  6.0 &  9.0 &  4.0 & 1.4 \\
        \datamixall{}     & \dataspmulti{} & full & Y & 2.5 &  3.6 & 10.1 &  3.1 & 0.8\\
        \datamixall{}     & \dataspmulti{} & full & N &  3.2 & 3.7  &  9.1 & 3.2 & 0.9 \\
        \bottomrule
    \end{tabular}
    }
\end{table}

\section{\textsc{\absd{}}: Simple Text-to-sound Generation and Infilling}\label{sec:ab_sd}

In this section, we present \textsc{\absd{}}, a model for text-guided generation of general sound. The task is also referred to as text-to-audio generation (TTA) in many prior works\citep{liu2023audioldm, huang2023make, kreuk2022audiogen}. It aims to generate general audios given a holistic text description. In contrast to text-to-speech synthesis, the text cannot be frame-wise aligned to audio. Furthermore, sound data only constitutes a small portion of the whole training data. Thus we investigate whether general audio pre-training is able to bring gains to generation of audios of specific domain, which we take sound generation as an example. While we focus on generation of sound events, the technique can similarly apply to other areas (e.g., music).

Most prior works~\cite{liu2023audioldm,ghosal2023text,liu2023audioldm2,huang2023make,yang2023diffsound} build the diffusion models upon a constrained latent space, commonly learned through autoencoding. Such strategy has shown to improve the data efficiency~\cite{rombach2021highresolution}. In this work, we adopt a different approach, which directly builds the flow matching network on auto-encoding based latent representation of \emph{raw waveforms}. Such methodology has been largely explored in the language model space~\cite{kreuk2022audiogen,copet2023simple,agostinelli2023musiclm}, which typically requires to build a billion-scale model to achieve comparable performance to the alternatives aforementioned. Here we show that by leveraging such simple strategy the flow matching models can achieve SOTA performance while being highly efficient (e.g., $>2$x smaller than~\cite{kreuk2022audiogen}).

\subsection{Method}
Similar to speech generation, we model the text-conditional sound distribution with flow matching. In contrast to learning phoneme encoding from scratch, we employ a pre-trained text encoder to map audio captions into word embeddings. Due to the lack of alignment between audio and text embedding, a cross-attention layer is applied in each transformer layer to allow the model attend to the whole text sequence in modeling the gradient distribution, similar to ~\citet{ghosal2023text,liu2023audioldm,liu2023audioldm2,kreuk2022audiogen}. 

Different from prior works in TTA such as AudioLDM~\citep{liu2023audioldm}, AudioLDM2~\citep{liu2023audioldm2}, Tango~\citep{ghosal2023text}, we do not rely on an off-the-shelf variational auto-encoder~\citep{kingma2014autoencoding} to map the low-level audio representation (mel spectrogram) into a latent space and model the distribution in the original embedding space directly. 
This streamlines the model architecture and reduces the necessity of introducing excessive trainable parameters during fine-tuning, thus bridging the gap between pre-training and fine-tuning.

Except for the cross-attention layers, all the remaining parameters are initialized based on the pre-trained  model introduced in \cref{sec:ab_ssl}. Similar to text-to-speech synthesis, parameter-efficient fine-tuning strategy like LoRA~\cite{hu2021lora} can be applied in text-to-audio generation. In practice, we observed fine-tuning the whole model leads to significantly better performance and thus choose to fine-tune the whole model by default (see \cref{sec:tta_ablation}).  

\boldparagraph{Multi-stage fine-tuning} 
Compared to transcripts for text-to-speech synthesis, high-quality audio captioning data are much more scarce. Typically, public audio captioning datasets include fewer than $1000$ hours of audios, which is orders of magnitude smaller than the speech datasets. On the other hand, the larger-scale sound data often contain noisy category labels and has distributional shift in the audio category~\citep{kim2019audiocaps}. To mitigate this issue, we divide the fine-tuning process into two stages, which is based on low-quality (e.g., tags) and high-quality (e.g., human written captions) audio descriptions respectively. Weights of the first model are used to initialize the subsequent model. We argue the labeled data used in first stage, despite its noisy nature, is helpful for learning the text conditional distribution (see \cref{sec:tta_ablation}).

\subsection{Tasks and Evaluation}\label{sec:ab_sd_task}
We consider the following two sound generation tasks: text-to-sound (TTA) generation and text-guided audio infilling (TAI). We use AudioCaps test set~\citep{kim2019audiocaps}, a standard benchmark for sound generation~\citep{kreuk2022audiogen,liu2023audioldm,liu2023audioldm2,yang2023uniaudio,lee2023voiceldm,ghosal2023text}, to evaluate all models. For TTA, the model is evaluated standard Frechet Audio Distance (FAD)~\citep{Kilgour2019FrchetAD}, Frechet Distance (FD) and KL divergence (KLD) based on the pre-trained audio event tagger PANN~\citep{Kong2019PANNsLP}, and Inception score (IS)~\citep{Salimans2016ImprovedTF}. 
FAD and FD measure distribution-level similarity between reference samples and generated samples. KLD is an instance level metric computing the divergence of the acoustic event posterior between the reference and the generated sample for a given description. IS measures specificity and coverage for a set of samples without requiring references, which assigns a higher score if instance posteriors have low entropy and marginal posterior has high entropy.
The metrics are implemented following the \texttt{audioldm\_eval} toolkit.\footnote{\url{https://github.com/haoheliu/audioldm_eval}}. In addition, we calculate the similarity between generated audio and text description using the CLAP model~\cite{wu2023large}~\footnote{We use the \textit{630k-best} checkpoint of \url{https://github.com/LAION-AI/CLAP}}. 

In TAI, the model is conditioned on $p\%$ of the ground-truth audio as context to infill the remaining $(100-p)\%$, in addition to the text description of the whole audio. In particular, $p$ is set to be 30 and the middle $70\%$ are the region to fill in. In addition to the metrics for TTA, we further measure the similarity to the reference audio (\emph{CLAP-aa}), which is the cosine similarity between CLAP embeddings of the generated and reference audio.

In addition to the objective metrics aforementioned, we also conduct subjective evaluation to evaluate two main aspects of the generated audio: overall naturalness (OVL) and relevance to text input (REL), similar to~\cite{kreuk2022audiogen,liu2023audioldm}. 
For these two metrics, raters were asked to rate the perceptual quality and the match between audio and text of the audio samples in a range between 1 and 5 similar to MOS. Based on the evaluation protocol~\cite{kreuk2022audiogen}, the subjective evaluation is done on 100 randomly sampled files from AudioCaps test set. Each sample is evaluated by 5 annotators from professional annotation service. We list the annotation interface in \cref{sec:app-mos-eval}.

\subsection{Experimental Setup}\label{sec:ab_sd_exp}

\boldparagraph{Data} 
To train \textsc{\absd{}}, we use about 6K hours of audio data, among which $\sim 150$ hours are captioned audios (\datasdcap{}) and the remaining ones only consist of audio tags (\datasdtag{}). During the first-stage fine-tuning, the whole dataset is used while only the captioning data are used in the second stage. To tackle the ontology of audio tags, we concatenate the tags of different levels as the pseudo-caption of the audio. See \cref{tab:tta-caption-dataset} for example audio description in these two sources.

\begin{table}[ht]
    \centering
    \caption{Examples of audio descriptions in tag-based and caption-based datasets (Note: the two columns of each row are unaligned.)}
    \label{tab:tta-caption-dataset}
    \begin{tabular}{cc}
        \toprule
        Tag-based description & Caption-based description \\
        \midrule
       Animal & A woman talks nearby as water pours \\
        Drill & Multiple clanging and clanking sounds \\
        Fill, Liquid  & The sizzling of food while a dish is clanking \\
        Bell, Hall, Room, Inside, Large  & a motorboat cruises along, and a man talks \\
        \begin{tabular}{c}
             Wolves, Domestic, Animal, Canidae, Dogs, Pets  \\
             Bark, Bow-wow, Animals, Growling \\
        \end{tabular}  & \begin{tabular}{c}
             The wind is blowing, insects are \\
             singing, and rustling occurs 
        \end{tabular}  \\
        \bottomrule
    \end{tabular}
\end{table}

\boldparagraph{Implementation Details} We use T5-base~\citep{raffel2020exploring} to map the text description into embeddings. Each cross-attention layer has 16 heads and its implementation remains same as the self-attention layers except that keys and values are text embeddings.  
The time-step embedding is added to the T5 embedding before being attended to. In the first stage, we fine-tune the model for 200K updates with an effective batch size of 720K frames. During the second stage, we further fine-tune the model for 100K updates with an effective batch size 240K frames. For both stages, the learning rate and gradient clipping are set to 0.0002 and 0.2 respectively. For inference, we use \texttt{dopri5} solver with absolute and relative tolerance of $10^{-5}$ as the default option. The classifier-free guidance weight is tuned between 0 and 5 and we found setting it to 1 leads to the best result. 
For each text prompt, we generate 32 random samples and select the one with the highest CLAP similarity to the text prompt. For audio infilling, the masked audio is always kept for conditioning and only the text description is optionally dropped for classifier free guidance.

\boldparagraph{Baselines} We compare \absd{} against models from the faimily of AudioLDM2~\cite{liu2023audioldm2} and TANGO~\cite{ghosal2023text}, which stand as current SOTA approaches for general audio generation~\cite{liu2023audioldm2}.

\subsection{Main Results}

\boldparagraph{Text-To-Audio} 
\cref{tab:tta-sound-main} compares our model to prior audio audio generation models in TTA. \textsc{\absd{}} consistently outperforms all prior works in both objective and subjective evaluation by a large margin, though it is significantly more parameter efficient. It is also worth noting compared to many approaches listed in \cref{tab:tta-sound-main}, the sound training data we used is also fewer. This further reveals the effect of general domain pre-training for sound generation. 

\boldparagraph{Text-To-Audio Infilling} 
\cref{tab:tta-sound-infill} shows the the performance of \textsc{\absd{}} on TAI, as well as its comparison to prior works. Our model outperforms prior works by a large margin as well on this task. Compared to TAI, we noticed a mixing result according to different metrics. Noticably, the trend on FAD and KLD is not consistently, as in the comparison between TTA and TAI. This can be related to the sensitivity of metrics. On the other hand, the similarity between the generation and reference is greatly increased (CLAP-aa: 0.61$\rightarrow$0.77) when the context is fed into the model, which suggests the improvement of coherence to the original audio when context is employed.

\begin{table}[ht]
    \centering
    \caption{Text-to-audio generation results on AudioCaps evaluation set. Baselines are evaluated based on the respective official repos. Subjective scores are computed based on 95\% confidence interval.}
    \label{tab:tta-sound-main}
    \resizebox{\linewidth}{!}{
    \begin{tabular}{c|ccccc|cc}
        \toprule
                    & \multicolumn{5}{c|}{objective} & \multicolumn{2}{c}{subjective}\\
                    & FAD $\downarrow$ & FD $\downarrow$ & KLD $\downarrow$ & IS $\uparrow$ & CLAP $\uparrow$ & OVL $\uparrow$ & REL $\uparrow$\\
        \midrule
        Ground-truth & -	& -&	- &	13.28 &	0.49 & 3.36\ci{0.18} & 3.86\ci{0.18} \\
        \midrule
        AudioLDM-L-Full~\cite{liu2023audioldm} & 3.37 &	28.76 &	1.66 &	8.72 &	0.43 & 2.48\ci{0.14} & 3.20\ci{0.18} \\ 
        AudioLDM 2-Full~\cite{liu2023audioldm2} & 1.76 &	32.12 &	1.71 &	8.56 &	0.43 & 2.90\ci{ 0.16} & 2.98\ci{ 0.19} \\
        AudioLDM 2-Full-Large~\cite{liu2023audioldm2} & 1.89 &	33.28 &	1.60 &	8.55 &	0.45 & 2.90\ci{ 0.16}& 3.13\ci{0.17}\\
        TANGO~\cite{ghosal2023text} & 1.57 &	23.78 &	1.37 &	8.30 &	0.51 & 3.10\ci{0.14} & 3.51\ci{0.16}\\
        TANGO-full-FT~\cite{ghosal2023text} & 2.19 &	18.47 &	1.20 &	8.80 &	0.56 & 3.04\ci{0.13} & 3.78\ci{ 0.15} \\
        \textsc{\absd{}} & \textbf{0.77} &	\textbf{8.30} &	\textbf{1.15} &	\textbf{12.70} &	\textbf{0.71} & \textbf{3.43\ci{ 0.15}} & \textbf{4.09\ci{0.15}} \\
        \bottomrule
    \end{tabular}
    }
\end{table}

\begin{table}[ht]
    \centering
    \caption{Text-to-audio infilling results on AudioCaps evaluation set. Baselines are evaluated based on the respective official repos. Subjective scores are computed based on 95\% confidence interval.}
    \label{tab:tta-sound-infill}
    \resizebox{\linewidth}{!}{
    \begin{tabular}{c|cccccc|cc}
        \toprule
                    & \multicolumn{6}{c|}{objective} & \multicolumn{2}{c}{subjective}\\
                    & FAD $\downarrow$ & FD $\downarrow$ & KLD $\downarrow$ & IS $\uparrow$ & CLAP $\uparrow$ & CLAP-aa $\uparrow$ & OVL $\uparrow$ & REL $\uparrow$\\
        \midrule
        Ground-truth & -	& -&	- &	13.28 &	0.49 & - & 3.13\ci{0.13} & 4.21\ci{0.15} \\
        \midrule
        AudioLDM-L-Full~\cite{liu2023audioldm} & 2.65 &	21.27	& 0.84 &	8.27 &	0.51 & 0.76 & 2.58\ci{0.12}& 3.58\ci{0.17} \\ 
        TANGO~\cite{ghosal2023text} & \textbf{1.25}	& 18.02 & 	0.78	& 8.53 & 0.53 & \textbf{0.78} & 2.75\ci{0.12} & 3.94\ci{0.15} \\ 
        TANGO-full-FT~\cite{ghosal2023text} & 1.86	& 15.00 &	0.71 &	8.95 &	0.56  & \textbf{0.78} & 2.79\ci{0.12}& 4.07\ci{0.14} \\
        \textsc{\absd{}} & 1.29 &	\textbf{7.19} &	\textbf{0.65} &	\textbf{12.05} &	\textbf{0.63} & 0.77 & {\textbf{2.95\ci{0.12}}} & {\textbf{4.20\ci{0.12}}} \\
        \bottomrule
    \end{tabular}
    }
\end{table}

\begin{figure}[h]
    \centering
    \includegraphics[width=0.75\linewidth]{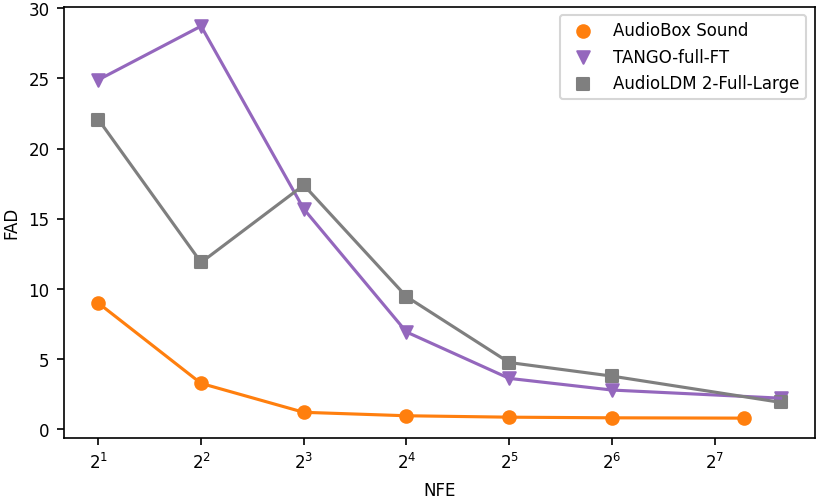}
    \caption{Quality-speed trade-off of \textsc{\absd{}}, Tango and AudioLDM2. NFE: Number of function evaluations.}
    \label{fig:sound_fad_nfe}
\end{figure}

\boldparagraph{Inference efficiency} In addition to quality metrics, we further show the quality-speed trade-off at inference time in \cref{fig:sound_fad_nfe}. 
Specifically, we vary the number of inference steps, which correspond to the step size in the ODE solver for our model and the number of DDIM steps in TANGO and AudioLDM2.
\textsc{\absd{}} achieves consistently higher quality (lower FAD) with the same number of inference steps compared to AudioLDM2 and Tango. This implies the better efficiency of the flow-matching approach \ab{} is based on, as is similarly demonstrated in~\cite{le2023voicebox}.

\subsection{Analysis and Ablation Study}\label{sec:tta_ablation}

\boldparagraph{Ablation Study} Here we conduct an ablation study showing the effect of different components of \textsc{\absd{}}. Specifically, we vary the following training strategies: training with \datasdcap{} only, training with \datasdtag{} and \datasdcap{}, training with the whole speech, music and sound datasets. 

As is shown in \cref{tab:tta-sound-ablation}, using a general pre-trained model boosts the performance by $\sim 20\%$  in FAD. Despite the discrepancy in task and data domain, generation of universal audios is a beneficial pretext task for text-to-sound generation. As music and speech constitutes a significant portion of our evaluation set, increasing the scale of these two modalities in pre-training provides additional benefits. 
Furthermore, the two-stage fine-tuning also consistently outperforms fine-tuning with \datasdcap{} only regardless of using a pre-trained model or not. The gain is mostly attributed to scaling up in-domain training data (i.e., sound only). Despite the labels being different, simply using audio tags can still enhance learning the mapping between the description of events and the actual audio. Finally, comparing the last two rows of \cref{tab:tta-sound-ablation} suggests reranking with CLAP model is an effective approach to improving the overall performance in both the audio quality (FAD: $0.91\rightarrow0.78$) and text-audio relatedness (CLAP score: $0.60\rightarrow0.71$).

\textbf{Fine-tuning strategy} We compare the two different fine-tuning strategies: LoRA vs. full model fine-tuning. For LoRA, we add LoRA adaptors described in \cref{sec:ab_sp} to self-attention layers. In contrast to full-tuning where the whole model is fine-tuned, only the adaptors and cross-attention layers will be updated during fine-tuning and all the remaining parts are frozen. LoRA fine-tuning
is on average $15\%$ to $30\%$ worse (relative)
than its full fine-tuning counterpart. The incorporation of cross-attention layers induces large architectural change to the model, which increases the necessity of fine-tuning the whole model.

\begin{table}[ht]
    \centering
    \caption{Ablation for sound generation on AudioCaps evaluation set. Tag: audio tagging data, Cap: captioning data. Note the results of this table are based on the midpoint solver with a step size of ${1}/{32}$ (equivalent to 64 NFE) for the purpose of inference speed-up.}
    \label{tab:tta-sound-ablation}
    \resizebox{\linewidth}{!}{
    \begin{tabular}{cccc|ccccc}
        \toprule
        PT (SSL) & FT-1 & FT-2 & w/ rerank & FAD $\downarrow$ & FD $\downarrow$ & KLD $\downarrow$ & IS $\uparrow$ & CLAP $\uparrow$ \\
        \midrule
        \xmark & - & \datasdcap{} & \cmark & 1.17 &	9.88 &	1.17 &	11.43 &	\textbf{0.71}  \\
        \xmark & \datasdtag{} + \datasdcap{} & - & \cmark & 1.61 &	13.16 &	1.34	& 10.17 &	0.67 \\
        \xmark & \datasdtag{} + \datasdcap{} & \datasdcap{} & \cmark & 0.97 &	8.70 &	1.17	& 12.19 &	\textbf{0.71} \\
        \cmark & - & \datasdcap{} & \cmark & 0.95 &	8.70 &	1.15 &	12.21 &	0.70 \\
        \cmark & \datasdtag{} + \datasdcap{} & \datasdcap{} & \xmark & 0.91 &	8.95 &	1.33 &	12.41 &	0.60 \\
        \cmark & \datasdtag{} + \datasdcap{} & \datasdcap{} & \cmark & \textbf{0.78} &	\textbf{8.31} &	\textbf{1.14} &	\textbf{12.62} &	\textbf{0.71} \\
        \bottomrule
    \end{tabular}
    }
\end{table}

\section{\textsc{\ab{}}: Toward Universal and Controllable Audio Generation}\label{sec:ab_uni}

In previous sections, we discussed speech and sound generation independently. This section presents \textsc{\ab{}}, a single model that can produce both speech and audio conditioned on text description or audio example. Fine-tuning our pre-trained model for this joint task enables natural language instruction to control the output speech attributes like perceived age, gender, quality on top of example-based control (ZS-TTS). Furthermore, training on wide variety of data enables simulating voices in different environments and accompanied by acoustic events such as birds chirping, applause. We further envision a scenario where the user would like to \textit{restyle} the given audio example with natural language instruction. For example, change the audio style to make it sound like it is recorded in a cathedral. This requires disentangled vocal style control using an additional utterance from the same speaker called voice prompt. 

We design \textsc{\ab{}} to enable speech and sound generation capabilities previously discussed in \cref{sec:ab_sp,sec:ab_sd}. Furthermore through voice prompt and description we also envision vocal style transfer to more complex acoustic scenes enabled through joint training.
Below we discuss in details speech caption and voice prompt modeling, data creation, and experiments.

\subsection{Data Creation}

\subsubsection{Speech Captions}\label{sec:ab_spcap}
We aim to bridge the gap between speech and sound datasets by supporting description-based control for speech generation. We consider both human annotations and automatically created captions

\boldparagraph{Automatic captions}
Given the lack of any dataset with fine-grained description for speech, we generate speech captions using a large language model (LLM) with speech attribute tags extracted either using existing metadata or use pseudo labels using classifiers. We extract the following attributes:
\begin{enumerate*}[label=(\arabic*)]
    \item age: 4 classes
    \item gender: 2 classes
    \item audio quality: 3 classes
    \item pitch: 3 classes
    \item speaking rate: 3 classes
    \item accent: open-vocabulary
    \item emotion: open-vocabulary
    \item environment: open-vocabulary
\end{enumerate*}
More details can be found in \cref{sec:app_sp_attr}.

Given the above attributes, we use the LLAMA2 7B model \cite{Touvron2023Llama2} to convert them into captions. To capture different writing styles, we prompt the model a style bank mimicking different characters with example writing samples. A few of them are listed below:

\begin{itemize}
    \item A young male adult voice, conveys anger and frustration. The audio, of normal quality, is recorded inside a small space. The person speaks with South Asia accent and a normal speaking pace.
    \item This young bloke's ticked off, audio's all good. He's in some small space and has a South Asian accent. Talks normal speed.
    \item Got this young dude who's mad, audio's decent. He's in a tight spot, has that South Asian accent, and talks at a chill pace.
    \item Young man is angry. Audio is okay, small place. Accent from South Asia. Speaks normal.
\end{itemize}

To further improve coverage over different environment and background sounds, for each utterance, we apply a random augmentation by convolving with a random room impulse responses (RIR) from a set of known environments and optionally add add a background noise from a set with known tags. 

We also generate the corresponding caption with updated environment and background noises using the LLAMA2 7B model. When adding any background noise to the utterance, we update the quality to ``low''. For utterances applied only RIR we update the quality to be ``normal'' if the original quality was ``studio''. 
We do not apply utterances with low audio quality since those may not be suited for RIR augmentations.

\boldparagraph{Human annotations}
We create human-based annotation to gather more fine-grained description and better alignment towards human hearing perception. We select a 500 hour subset of \dataspmulti{} described in \cref{sec:ab_sp_ablate}.

In the annotation guidelines, we ask the annotator to describe the perceived attribute such as: gender, age, accent, emotion, environment, tonal variation, speaking pace, pitch, emotion, audio quality, vocal style and any miscellaneous details from the speech utterances. In addition to this we also collect categories for the attributes. To ensure we get high quality description, we filter annotators in two stages. First, we keep annotators who successfully labeled pre-selected gold samples with high accuracy. We additionally use an LLM to automatically rate the quality annotations to ensure high quality detailed captions to complement our automatic caption above. More details on quality can be found in \cref{sec:app_caption_quality}. Here are some captions example curated by our human annotator: 

\begin{enumerate}
    \item A young woman with an American accent speaks in a higher pitched voice. She speaks at a normal pace with a bit of a muffled voice. She is outside in an urban area and cars can be heard passing by in the background. She has a happy and excited tone that is slightly melodious. The audio is of poor quality and dog barking can be heard at the end.
    \item A middle aged man with a mildly masculine voice seems to be outside in a rural or natural environment with a moderately background noise of birds singing. He seems to be in a neutral mood when show casing a house to some people. His voice is hoarse/rough speaking at a slow pace with an average voice pitch.
\end{enumerate}

\subsubsection{Voice Prompts}
\label{subsubsec:voice_prompts}

Natural language description alone allows user to control styles through describing attributes such as age, accent, emotion, pitch, and environment. However, a user maybe interested in synthesizing a specific vocal style and while changing other attributes such as quality, emotion, background. This requires disentangled control between the input voice sample and natural language text prompt. 

For each target utterance, we sample an additional utterance from the same speaker to serve as voice prompt during training. The voice prompt is selected such that it differs from the target utterance on one or more attribute such as emotion, environment, and speaking rate. This is to de-correlate the target and prompt on everything but vocal similarity. We additionally apply a random room impulse response and background noise augmentation to the voice prompt to increase robustness as well as further de-correlation.

Note that this is different from passing the audio as audio context (zero-shot TTS) where we expect the model to copy over emotion, environment and other background details as well. Here we would want the model to transfer only the vocal style from prompt and use the description for other details such as environment and emotions.

\subsection{Method}

\textsc{\ab{}} (\cref{fig:audiobox}) conditions on both transcript and masked audio features (same as \textsc{\absp{}}) and captions (same as \textsc{\absd{}}) for description conditional generation. To unify training, for sound inputs without transcript, we create a pseudo-transcript that contains ``<sound>'' tokens each of length 1 second filling the length of audio. We additionally condition on the another utterance from the same speaker (voice prompt). As described in \cref{subsubsec:voice_prompts}, the voice prompt is selected in a adversarial fashion to enable disentangled control. For audios with missing prompts, we feed a pseudo voice prompt of length $0.1$s filled with zeros. The voice prompt is embedded by a lightweight Transformer. We then concatenate the output with the caption description embedding for cross-attention. We randomly initialize the parameters for the cross-attention, description projection, and character embedding weights. All other parameters are initialized based on \textsc{\abssl{}} in \cref{sec:ab_ssl}. Similar to the sound model training in \cref{sec:ab_sd}, we use multi-stage fine-tuning as described next. 

\boldparagraph{Multi-stage fine-tuning} Except for the high quality $500$ hours of speech captions that we collect, the rest of our speech captions are generated using attribute tags and an LLM. Furthermore most of the datasets do not provide any meta-data further limiting the quality of the captions. To mitigate this issue we train our model in two stages. In the first stage we use all the captions for speech and audios. To avoid under-fitting on the audio events generation, we upsample the audio data such that the ratio of total speech and audio data in hours is about $3:1$. In the second stage, we initialize the model from first stage weights and only train on the high quality data that comprises $500$ hour of annotated speech captions and a few other datasets with emotion and accent metadata for rich LLM captions. We again upsample the audio data such that the ratio of total speech and audio data is about $2.6:1$.

\subsection{Task and Evaluation}
In our unified \textsc{\ab{}} model, the model is capable of new generation tasks such as description-guided TTS (transcript + description) and description-guided TTS with extra voice conditioning generation (transcript + description + voice prompt). Additionally, \textsc{\ab{}} also maintains generation capability from all prior section including: diverse speech sampling (transcript only), zero-shot TTS (transcript + context prompt) (see \cref{sec:ab_sp_task}), text-to-sound (TTA) generation (description only) and text-guided infilling (TAI, description + context prompt) (see \cref{sec:ab_sd_task}). In \cref{sec:app_task_inputs}, describe the tasks and inputs in detail. 

For all speech generation tasks, we measure the WER and similarity of vocal style if context or voice prompt is provided. In addition, for any generation task with description conditioning, we measure the similarity between description and generated audio with cosine similarity between CLAP text and audio embedding. 
For the description-guided TTS, in addition to objective metric, we also conduct subjective evaluation to assess the QMOS and REL. Below, we provide details on the CLAP model used for speech evaluation.

\subsubsection{Joint-CLAP similarity} 
In terms of tasks, generating speech conditioned on text descriptions is similar to description-guided sound generation (TTA). As is common in TTA, we also employ the text-to-audio similarity to measure how well the generated audio matches the description. However, unlike TTA scenario, joint text-audio embedding models such as CLAP~\cite{wu2023large} cannot be straightforwardly applied to the speech domain. Existing CLAP models are trained with coarse description about speech, such as "a person speaking". The model is unable to distinguish fine-grained speaking styles like accent or emotion. Although there exist public CLAP models which are trained with speech data, most of them are trained with (speech, transcript) pairs which is orthogonal to the text description. Thus, for the purpose of evaluating description-conditioned speech generative models, we propose \emph{Joint-CLAP} model, which is designed for both description-based speech and audio evaluation.

\textbf{Training} Similar to CLAP~\cite{wu2023large}, Joint-CLAP consists of an audio and text branch, each responsible for encoding audio waveforms and the natural language sentences respectively. Given a speech-text pair $(x^a, x^t)$, the audio and text branch $f_{a}$ and $f_{t}$ encodes it into the embedding pair $(e^a, e^t)$: $e^a=f_a(x^a)$, $e^t=f_t(x^t)$.  We use the same contrastive loss for model training following~\cite{wu2023large,Radford2021LearningTV}, where $\tau$ is a learnable parameter.

\begin{equation}
\label{eq:speech-clap-loss}
    L=\frac{1}{2N}\displaystyle\sum_{i=1}^{N}(\log\frac{\exp{(e_i^a\cdot e_i^t}/\tau)}{\sum_{j=1}^N\exp{(e_i^a\cdot e_j^t/\tau)}}+\log\frac{\exp{(e_i^t\cdot e_i^a}/\tau)}{\sum_{j=1}^N\exp{(e_i^t\cdot e_j^a/\tau)}})
\end{equation}

In practice, we use pre-trained RoBERTa~\cite{Liu2019RoBERTaAR} as the text encoder $f_t$. In contrast to CLAP, which uses pretrained audio taggers (e.g., HSTAT~\cite{chen2022htsat}) for audio encoding, here we use WavLM~\cite{Chen2021WavLMLS} as the backbone for encoding. Self-supervised speech models can better capture detailed information (e.g., speaking style) than general audio classifiers. Both RoBERTa and WavLM encoders are fine-tuned in model training.

\textbf{Data} The training data of Speech-CLAP consists of \datasdtag{}, \datasdcap{}, and 2K hours of speech datasets including both human and automatic captions. The training set includes both speech and non-speech data in order to equip the model the discriminative capabilities for speaking with environmental sound use cases (e.g., \emph{a man speaks as birds chirp and dogs bark}). The speech portion is a subset of the captioned speech described in \cref{sec:ab_spcap}, which are selected to balance the ratio of human annotated and LLM-augmented captions. The model is evaluated on the evaluation sets of the sound and speech subset respectively.  

\textbf{Implementation Details} For audio and text encoder, we use WavLM-base+ and RoBERTa base respectively. Using alternative speech encoders within the same family such as WavLM-large brings similar results. The audio and text embeddings are normalized before calculating the loss (\cref{eq:speech-clap-loss}). The model is trained using Adam optimizer~\cite{Kingma2014AdamAM} with a learning rate of $5e-5$. We use 64 volta32 GPUs with a batch size of 75 per GPU for 200K updates. For training stability, the gradient is clipped to 10 by norm and raw floating point precision is used without any quantization. We track the recall (A2T@10) on the validation set at the end of each epoch and select the model checkpoint with the highest value. 

\textbf{Retrieval Performance} We compare Joint-CLAP to the original CLAPs proposed by~\cite{wu2023large}, measuring the text-to-audio and audio-to-text retrieval performance. Specifically, we take two public CLAP models trained general audios: \emph{CLAP (general audio)}~\footnote{\url{https://huggingface.co/lukewys/laion\_clap/blob/main/630k-best.pt}}, and general audios plus speech: \emph{CLAP (w/ speech)}~\footnote{\url{https://huggingface.co/lukewys/laion_clap/blob/main/music_speech_audioset_epoch_15_esc_89.98.pt}} 
Per retrieval task, we report the recall under three thresholds: 1, 5 and 10. As is shown in \cref{tab:ab_speech_clap}, public CLAPs, regardless of whether speech data are utilized or not, achieves significantly lower performance on speech retrieval based text descriptions, with $\sim 30$x performance degradation compared to the sound benchmark. This might be due to the naturally larger ambiguity in the task, where description of speech may exhibit higher variance. For instance, different people may have varying opinions on what constitutes fast speaking versus slow speaking.
In spite of such ambiguity, Joint-CLAP still significantly improves the retrieval performance under the same setting (T2A@10 on speech: $2.29\rightarrow 22.01$), while maintaining the performance for general audios (T2A@10 on sound: $63.64\rightarrow 67.64$). The gain is attributed to fine-tuning with speech-specific datasets and using a high-performing speech encoder. To further ablate this effect, we trained a CLAP model without altering the model architecture using in-domain speech data. The retrieval performance is considerably lower than the WavLM-based Joint-CLAP (e.g., T2A@10 on speech: $12.01$ vs. $22.01$). 

\begin{table}[htb]
    \centering
    \caption{Comparison between Speech-CLAP and public CLAP models on retrieval performance in sound and speech. }
    \label{tab:ab_speech_clap}
    \begin{tabular}{c|cccccc}
        \toprule
        & \multicolumn{6}{c}{Speech} \\
        & \multicolumn{3}{c}{Text$\rightarrow$Audio} & \multicolumn{3}{c}{Audio$\rightarrow$Text} \\
        & R@1 & R@5 & R@10 & R@1 & R@5 & R@10 \\
        \midrule
        \begin{tabular}{c}
           CLAP (general audio)~\cite{wu2023large}
        \end{tabular} & 0.36 &	1.29 &	2.29 &	0.64 &	2.26 &	3.55\\
        \begin{tabular}{c}
           CLAP (w/ speech)~\cite{wu2023large}
        \end{tabular} & 0.82 &	2.42 &	3.37 &	0.51 &	1.90 &	2.60 \\
        Speech-CLAP & \textbf{7.10} &	\textbf{16.30} &	\textbf{22.01} &	\textbf{5.96} &	\textbf{16.07} &	\textbf{22.34} \\
        \midrule\midrule
        & \multicolumn{6}{c}{Sound} \\
        & \multicolumn{3}{c}{Text$\rightarrow$Audio} & \multicolumn{3}{c}{Audio$\rightarrow$Text} \\
        & R@1 & R@5 & R@10 & R@1 & R@5 & R@10 \\
        \midrule
        \begin{tabular}{c}
           CLAP (general audio)~\cite{wu2023large}
        \end{tabular} &	11.03 &	45.33 &	63.64 &	9.45 &	44.36 &	61.70 \\
        \begin{tabular}{c}
           CLAP (w/ speech)~\cite{wu2023large}
        \end{tabular} & 11.15 &	42.42 &	60.36 &	9.70 &	43.15 &	59.03 \\
        Speech-CLAP &	\textbf{13.33} &	\textbf{51.88} &	\textbf{67.64} &	\textbf{11.27} &	\textbf{47.27} &	\textbf{64.48} \\
        \bottomrule
    \end{tabular}
\end{table}

\begin{figure}[h]
    \centering
    \includegraphics[width=0.75\linewidth]{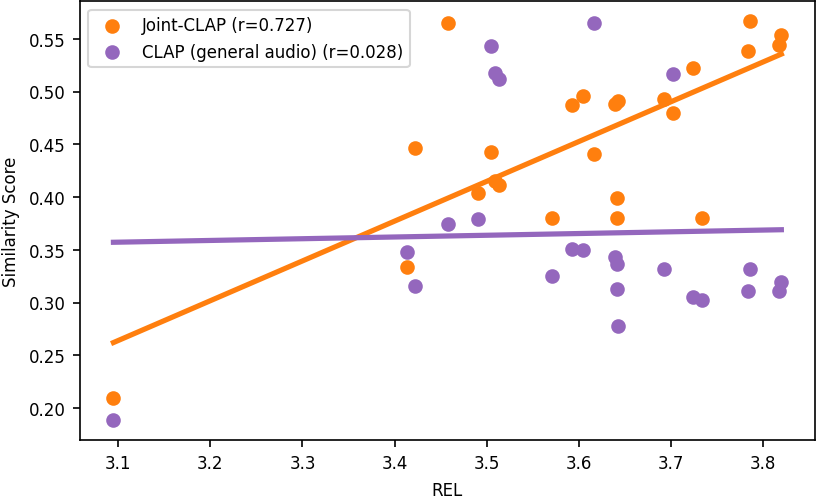}
    \caption{Correlation between text-audio similarity and REL score in different CLAP models. $r$: Pearson correlation coefficient.}
    \label{fig:joint_clap_corr}
\end{figure}

\textbf{Correlation between Joint-CLAP scores and human opionion scores} In practice, we also notice the Joint-CLAP model is more closely correlated to human-perceived text-audio similarity, as opposed to the public CLAP model (see \cref{fig:joint_clap_corr}). Specifically, we take six \textsc{\ab{}} models of varying performance and run subjective evaluation with these models on the four evaluation sets. As is shown in \cref{fig:joint_clap_corr}, the Pearson correlation coefficient between the text-audio similarity and REL score is increased from 0.028 to 0.727 with a joint CLAP model, suggesting that its text-audio similarity score is a reliable metric for evaluating description-controlled speech generation.

\subsection{Experimental Setup}

\boldparagraph{Training data} We train unified \textsc{\ab} with a combination of (1) English speech dataset (SP-Multi-100K, see \cref{sec:ab_sp_exp}) with additional text description and voice prompt for each corresponding utterances and (2) sound dataset with text description or tags (SD-TAG-6K and SD-CAP-150, see \cref{sec:ab_sd_exp}). In both cases, each description is either generated from an LLM, or annotated by humans. We employ two-stage fine-tuning to improve our model fidelity and quality. In the first stage fine-tuning, we incorporate all speech (SP-Multi-100K) and sound (SD-TAG-6K and SD-CAP-150) datasets into our training dataset. In the second stage fine-tuning, we use a subset of our first-stage fine-tuning dataset comprised of higher quality dataset with total about 2,310 hours.

\boldparagraph{Implementation details}
Unified \textsc{\ab} model takes four different inputs: 1) frame-aligned transcript, 2) description, 3) voice prompts, and 4) context prompt (masked audio features). First, we first embed the input character sequence in frame-aligned transcript to $128$ dimension features. The embedded sequence is then projected using a linear layer and added to the projected masked audio features as input to the Transformer. Next, we use T5-base to extract 512-dimension continuous embedding from the description. The parameters of T5-base are kept frozen during training. We add a trainable linear layer to project the output from $512$-dimensions to match the Transformer embedding dimensions ($1024$). For the voice prompts, we first extract dense features using the same Encodec model described in \cref{sec:ab_ssl}. These features are then input to a $3$-layered Transformer model with $1024$ embedding dimensions, $16$ attention heads, and a feed-forward dimension of $4096$. We then concatenate the time-step embedding, voice prompt encoder output, and description embedding which form the input for cross-attention. 

During training, we randomly drop voice prompt, captions, and context with the probabilities specified in \cref{tab:voice_prompt_drop_probs}:
\begin{table}[ht]
    \centering
    \caption{Drop-out probabilities for context (ctx), voice prompt (vp), and caption (cap). ``F'' (false) / ``T'' (true) refers whether the input is used.}
    \label{tab:voice_prompt_drop_probs}
    \begin{tabular}{ccccc}
        \toprule
        \multicolumn{4}{c}{Hyper-parameters} \\
        P(vp=F) & p(ctx=F | vp=T) & P(ctx=F | vp=F) & p(cap=F) \\
        \midrule
        0.5	& 0.7 & 0.5	& 0.3 \\
        \bottomrule
    \end{tabular}
\end{table}

These probabilities are designed with specific use cases discussed previously.  Note that zero-shot TTS requires the model to copy each and every attribute from the audio prompt while restylization requires model to maintain high similarity of vocal style while discarding emotion, environment and other attributes. This requires us to distinguish the context from the voice prompt.

Setting the dropout probabilities as defined in \cref{tab:voice_prompt_drop_probs} lead to the joint probabilities presented in \cref{tab:joint_voice_prompt_drop_probs}. The joint probabilities correspond to each of the use case that the model can support. Note that the generative pre-training already tunes model for \textit{ZS-TTS} and \textit{diverse speech sampling} applications. Therefore, we select the hyper-parameters to bias the model towards \textit{description-guided TTS} with and without vocal conditioning.

\begin{table}[ht]
    \centering
    \caption{Derived joint probabilities for context, voice prompt, and caption for different use cases.}
    \label{tab:joint_voice_prompt_drop_probs}
    \resizebox{\linewidth}{!}{
    \begin{tabular}{cccccc}
        \toprule
        \multicolumn{4}{c}{Hyper-parameters} \\
        ZS-TTS & Description-TTS w/ vocal & Description-TTS & Sampling \\
        P(ctx=T, vp=F, cap=F) & P(ctx=F, vp=T, cap=T) & P(ctx=F, vp=F, cap=T) & P(ctx=F, vp=F, cap=F) \\
        \midrule
        0.075	& 0.245 & 0.175	& 0.075 \\
        \bottomrule
    \end{tabular}
    }
\end{table}

In the first stage fine-tuning, we fine-tune all parameters for a maximum of 600K updates with 32 A100-80GB GPUs. We stopped training after 350K steps as we didnot find any gains in model performance beyond this. In the second stage, we further fine-tune our model parameter with LoRA fine-tuning on the self-attention parameters with $r=64$ and cross attention input projection layers for 100K updates with 16 A100-80GB GPUs.

For the unified \textsc{\ab{}} duration model, we use both transcript and the description text as the input. We use 12 Transformers decoder layer with 8 heads, 768/2048 embedding/FFN dimensions self-attention and cross-attention layer to attend the description embedding. We use 40 dimension for the character embedding. During training, we set the description embedding drop probability 0.3. The model trained with 600K updates with flow-matching loss with 8 A100-80GB GPUS. For evaluation, we use the checkpoint at 200K steps.

\boldparagraph{Evaluation data} 
We measure the effectiveness of description-guided TTS and description-guided TTS with vocal prompts on the following test sets. 
First, we annotate a set of 1,946 recordings sampled from diverse sources, including LibriTTS~\citep{zen2019libritts}, Common Voice~\citep{Ardila2019CommonVA}, Switchboard~\citep{godfrey1992switchboard}, Fisher~\citep{cieri2005fishers, cieri2005fishert}, Spotify~\citep{clifton2020spotify}, AudioSet~\citep{gemmeke2017audio}, Expresso~\citep{nguyen2023expresso} in order to evaluate the ability to generalize. This set is denoted as SpCap (SC).
The second set is AC-filtered (AC-filt)~\citep{lee2023voiceldm} with 825 utterances. It constructed from AudioCaps test set by transcribing and keeping samples with reliable ASR transcriptions. 

The third one is the Expresso test set (Expr) with 999 utterances. 
Finally, the fourth one contains utterances from the internal Accent set. We apply randomly sampled RIR and noise augmentation to construct this set and denote it as ``Accent+'' (500 utterances). 
Expr and Accent+ use speech captions derived from LLM using the available attributes. For Accent+, we additionally pass the environment and background noises tags to the LLM to incorporate the information into generated captions. 
Together these sets cover a wide variety of acoustic events, emotions, accents, environments, and vocal styles.

To evaluate description-based TTS with vocal prompt, we use Expr and Accent+ datasets and select another utterance from the same speaker. The prompt is selected such that is different from the target utterance on either emotion or speaking style (enunciated, whisper, etc).  Furthermore, we also compare against \textsc{\absd{}} and \textsc{\absp{}} on speech and sound applications using the evaluation sets described in \cref{sec:ab_sp,sec:ab_sd} respectively.

\boldparagraph{Inference} We use duration model described in this section with averaging over $5$ samples. For description-guided TTS (with or without voice prompt), we additionally sample a silence duration of between $0$ and $3$ seconds and pad it to both ends. We find this generates audios that are coherent with the description particularly when they also mention acoustic events. 
For example: a man speaks and car passes by while a dog is barking. However, this can cause model to hallucinate sounds when there are no acoustic events described. 
To cover all scenarios involving description-guided TTS, we generate $N=8$ samples with stochastic silence padding and then output the best sample based on clap re-ranking using the joint model. We use a guidance weight of $0.75$ for the description-guided TTS (with/without voice prompt) applications.

For sound only generation, we always generate $10$s long audios with pseudo-transcripts using a guidance weight of $1.33$.  We use clap reranking with $N=16$ samples using the sound clap model. 
For zero-shot in-context TTS applications, we trim the end-silences similar to the \textsc{\absp{}} model and use a guidance weight of $1.0$. Given that this application doesn't involve any descriptions, we do not use clap re-ranking.
Unless specified, both acoustic and duration \textsc{\ab} models use the midpoint solver with a step size of $1/32$, which invokes the function being integrated 64 times. When using classifier free guidance the model does 2 forward passes, leading to a total of 128 calls to the model forward pass.

\subsection{Main Results}
In this section, we investigate the effectiveness of the unified \textsc{\ab{}} model on a number of use cases. We first compare the description-guided TTS with and without voice prompt in \cref{tab:ab_desc_tts,tab:ab_desc_vp_tts} respectively. For this task, we compare with VoiceLDM \cite{lee2023voiceldm} and AudioLDM2 \cite{liu2023audioldm2} models as baselines. Next, in \cref{tab:ab_inctx_tts_spk} we evaluate how well \textsc{\ab{}} performs speech tasks as compared to non-description speech only model, \textsc{\absp{}}. Finally, in \cref{tab:ab_tta} we compare against the sound-only \textsc{\absd{}} model on the TTA task.

\ava{}

\subsubsection{Description-based control for speech generation}

\cref{tab:ab_desc_tts} compares \textsc{\ab{}} with VoiceLDM \cite{lee2023voiceldm} and AudioLDM2 \cite{liu2023audioldm2} models on description-guided TTS and description-guided TTS with voice prompt  (voice restylization) tasks.  We find that \textsc{\ab{}} outperforms both baselines on all datasets and metrics. In particular,  \textsc{\ab{}} is able to consistently generate audios for rich descriptions in SC, background events (AC-filt), expressive audios (Expr), and accented audios with diverse backgrounds (Accent+).

\begin{table}[ht]
    \centering
    \caption{Description-based control for speech generation. \textsc{\ab{}} outperforms both AudioLDM2 and VoiceLDM on all datasets and metrics. VoiceLDM and AudioLDM2 models struggle in particular of Expr and Accent+ datasets with expressive audios.}
    \label{tab:ab_desc_tts}
    \resizebox{\linewidth}{!}{
    \begin{tabular}{c|cccc|cccc}
        \toprule
        & \multicolumn{4}{c}{JointCLAP $\uparrow$} & \multicolumn{4}{c}{WER (\%) $\downarrow$} \\
        & SC & AC-filt & Expr & Accent+ & SC & AC-filt & Expr & Accent+ \\
        \midrule
        ground truth & 0.403 & 0.479 & 0.548 & 0.561 & 8.4 & 23.5 & 5.8 & 13.5 \\
        \midrule
        VoiceLDM & 0.245 & 0.449 & 0.060 & 0.235 & 8.0 & 6.8 & 5.3 & 4.4 \\ 
        AudioLDM2-SP & 0.241 & 0.225 & 0.066 & 0.110 & 32.5 & 26.3 & 33.8 & 23.9 \\
        \textsc{\ab{}} &  \bf{0.430} & \bf{0.489} & \bf{0.387} &  \bf{0.596} & \bf{7.2} &  \bf{5.2} &  \bf{4.5}   & \bf{2.6} \\
        \midrule\midrule
        & \multicolumn{4}{c}{QMOS $\uparrow$} & \multicolumn{4}{c}{REL $\uparrow$} \\
        & SC & AC-filt & Expr & Accent+ & SC & AC-filt & Expr & Accent+ \\
        \midrule

        ground truth &  {3.60\ci{0.11}} & {3.25\ci{0.14}} & {4.00\ci{0.09}} & {3.24\ci{0.13}} &  {3.66\ci{0.10}} & {3.86\ci{0.12}} & {4.01\ci{0.10}} & {3.51\ci{0.11}} \\
        
        \midrule
        
        VoiceLDM    &  {3.01\ci{0.10}} & {2.95\ci{0.13}} & {2.92\ci{0.12}} & {2.87\ci{0.12}} &  {2.90\ci{0.10}} & {3.08\ci{0.14}} & {2.78\ci{0.11}} & {3.2\ci{0.11}}       \\
        AudioLDM2-SP  &    {2.19\ci{0.11}} & {2.17\ci{0.12}} & {2.47\ci{0.11}} & {2.25\ci{0.10}} & {2.37\ci{0.11}} & {2.11\ci{0.12}} & {2.48\ci{0.11}} & {2.22\ci{0.10}}      \\
         \textsc{\ab{}}   &   \bf{3.58\ci{0.10}} & \bf{3.38\ci{0.12}} & \bf{3.82\ci{0.09}} & \bf{3.54\ci{0.12}} & \bf{3.74\ci{0.09}} & \bf{3.61\ci{0.12}} & \bf{3.94\ci{0.11}} & \bf{3.61\ci{0.10}}     \\
         
        \bottomrule
    \end{tabular}
    }
\end{table}

We also note that AudioLDM2 and VoiceLDM struggle in particular on expressive datasets (Expr and Accent+). In particular, we find that utterances generated by AudioLDM2 and VoiceLDM models are significantly worse than the ground truth especially in complicated scenarios involving description of both speech, environment (cathedral), and background sounds. This results in worse scores on the Accent+ dataset. Furthermore, Expr test set contains voices exploring expressive styles like enunciation, whispering, non-binary gender which is where AudioLDM2 and VoiceLDM struggle. We hypothesize this could be because they are out-of-distribution cases w.r.t training. Both VoiceLDM and AudioLDM2 model tend to struggle on such utterances leading to low scores on objective metrics.

{Our subjective evaluations also align with the objective metrics where we find the the \textsc{\ab} model significantly outperforms the baselines in particular to similarity to the description. The worse scores on Accent+ and Expr dataset for AudioLDM2 and VoiceLDM model further confirms our own observations.}

In \cref{tab:ab_desc_vp_tts}, we present the results for description-guided TTS with voice prompt. VoiceLDM and AudioLDM2 model do not simultaneously support conditioning based on vocal and text descriptions for a transcript. Towards our best effort comparison, we combine the CLAP embedding for the audio vocal prompt and the textual description by averaging them and use it as a conditioning input. We find that \textsc{\ab{}} outperforms both baselines. We also notice that in the absence of voice-prompt, the speaker similarity of \textsc{\ab{}} is greatly reduced as the description cannot capture all aspects of voice.  The subjective evaluations aligns with the objective metrics both for description and generated audio similarity and speaker similarity. We find that the voice prompt greatly improves the speaker similarity while matching the descriptions.

\begin{table}[ht]
    \centering
    \caption{Description-based control with extra voice conditioning for speech generation}
    \label{tab:ab_desc_vp_tts}
    \resizebox{\linewidth}{!}{
    \begin{tabular}{cc|cc|cc|cc}
        \toprule
        \multicolumn{8}{c}{Comparing on objective metrics.} \\
        \midrule 
        Model & Voice cond.
        & \multicolumn{2}{c}{JointCLAP $\uparrow$} & \multicolumn{2}{c}{Sim-o $\uparrow$} & \multicolumn{2}{c}{WER $\downarrow$} \\
        & & Expr & Accent+ & Expr & Accent+ & Expr & Accent+ \\
        \midrule
        ground truth  & n/a  & 0.548 & 0.561 & 0.395 & 0.526 & 5.8 & 13.5 \\
        \midrule
        VoiceLDM      & avg. CLAP & 0.093 & 0.204 & 0.115 & 0.076 & 4.8 & 3.9\\
        AudioLDM2-SP  & avg. CLAP & 0.067 & 0.118 & 0.045 & 0.089 & 34.6 & 30.2\\
        \textsc{\ab{}}      & No  & 0.387	& \bf{0.596}	& 0.181	& 0.141	& \bf{4.5} &	\bf{2.6}  \\
        \textsc{\ab{}}      & Yes  & \bf{0.480}	& 0.593	& \bf{0.377}	& \bf{0.344}	& 7.7	& 2.8 \\
        \midrule
        \multicolumn{8}{c}{Comparing on subjective metrics for Speaker similarity, quality and description aspects} \\
        \midrule
        Model & Voice cond.
        & \multicolumn{2}{c}{QMOS $\uparrow$} & \multicolumn{2}{c}{REL $\uparrow$} & \multicolumn{2}{c}{Speaker Similarity MOS $\uparrow$} \\ 
        && Expr & Accent+  & Expr & Accent+  & Expr & Accent+  \\ 
        \midrule

        ground truth & n/a & {4.0\ci{0.09}} & {3.24\ci{0.13}} & {4.01\ci{0.1}} & {3.51\ci{0.11}} & {3.38\ci{0.11}} & {3.27\ci{0.10}} \\
        \midrule
        
        \textsc{\ab{}}      & No  &  {3.82\ci{0.09}} & {3.54\ci{0.12}} & {3.94\ci{0.11}} & \bf{3.61\ci{0.1}}  & {3.02\ci{0.12}} & {3.03\ci{0.10}}  \\
                \textsc{\ab{}}      & Yes  &  \bf{3.86\ci{0.09}} & \bf{3.58\ci{0.12}} & \bf{3.99\ci{0.11}} & {3.57\ci{0.11}} & {\bf{3.36\ci{0.11}}} & {\bf{3.24\ci{0.11}}}  \\
        \bottomrule
    \end{tabular}
    }
    
\end{table}

\subsubsection{Comparison to \textsc{\absp} and \textsc{\absd}}
\cref{tab:ab_inctx_tts_spk} compares the unified \textsc{\ab{}} and speech only \textsc{\absp} models for zero-shot TTS on $5$ different datasets. We use the same duration model for both acoustic models for this task. We find that the unified \textsc{\ab} model gives higher speaker similarity but performs marginally worse on the word error rate. This is also confirmed by subjective evaluations where we find only minor differences between the \textsc{\ab} and \textsc{\absp} models.

In \cref{tab:ab_tta}, we present the results comparing the unified \textsc{\ab}  to the  \textsc{\absd}, VoiceLDM, and AudioLDM2 models on the task of TTA task as described in \cref{sec:ab_sd_task}. We find that  \textsc{\ab} significantly outperforms all baselines achieving the state-of-the-art performance for joint models and even outperforms sound only models such as TANGO. The  \textsc{\ab}  performs worse only to the  \textsc{\absd} model which specializes in sound generation. The subjective evaluations further confirm that both our \textsc{\ab} and \textsc{\absd} outperform all other baselines by a significant margin.

\begin{table}[ht]
    \centering
    \caption{Comparing \textsc{\ab} and \textsc{\absp} model for In-context TTS application. Both model use the same regression based duration model}
    \label{tab:ab_inctx_tts_spk}
    \resizebox{\linewidth}{!}{
    \begin{tabular}{c|cccccc|cccccc}
        \toprule
        \multicolumn{13}{c}{Style similarity and content correctness using objective metrics} \\
        \midrule
        & \multicolumn{6}{c}{Sim-o $\uparrow$} & \multicolumn{6}{c}{Word error rate (\%) $\downarrow$} \\
                          & LS    & CV    & SWBD  & Expr  & Accent & Avg & LS & CV & SWBD & Expr & Accent & Avg \\
        \midrule
        \textsc{\absp}    & \textbf{0.734} & 0.607	& 0.608	& 0.603	& 0.659	& 0.642 & \textbf{3.2} & 3.7  &  \textbf{9.1} & 3.2 & 0.9 & \textbf{4.0} \\
        \textsc{\ab} & 0.732 &  \textbf{0.624} &  \textbf{0.610} &  \textbf{0.643}  &  \textbf{0.674} &  \textbf{0.656}  & 4.8  & \textbf{3.0}  & 12.6  & \textbf{2.7}   & 0.9   & 4.8 \\        
        \midrule\midrule
        \multicolumn{13}{c}{Style similarity MOS subjective evaluation $\uparrow$} \\
        \midrule
        & \multicolumn{1}{p{0.1cm}}{}  & \multicolumn{2}{c}{LS} & \multicolumn{2}{c}{CV} & \multicolumn{2}{c}{SWBD} & \multicolumn{2}{c}{Expr} & \multicolumn{2}{c}{Accent} & \multicolumn{1}{p{0.1cm}}{}  \\
        \midrule

        \textsc{\absp} &  & \multicolumn{2}{c}{\textbf{3.88 $\pm$ 0.11}} & \multicolumn{2}{c}{3.77 $\pm$ 0.11} & \multicolumn{2}{c}{{3.63 $\pm$ 0.12}} & \multicolumn{2}{c}{3.85 $\pm$ 0.11} & \multicolumn{2}{c}{3.77 $\pm$ 0.11} & \\

        \textsc{\ab} &  &\multicolumn{2}{c}{{3.72 $\pm$ 0.11}} & \multicolumn{2}{c}{\textbf{4.03 $\pm$ 0.11}} & \multicolumn{2}{c}{\textbf{3.72 $\pm$ 0.12}} & \multicolumn{2}{c}{\textbf{4.01 $\pm$ 0.10}} & \multicolumn{2}{c}{\textbf{3.88 $\pm$ 0.11}} &  \\

        \midrule\midrule
        \multicolumn{13}{c}{Quality MOS subjective evaluation $\uparrow$} \\
        \midrule
        & \multicolumn{1}{p{0.1cm}}{}  & \multicolumn{2}{c}{LS} & \multicolumn{2}{c}{CV} & \multicolumn{2}{c}{SWBD} & \multicolumn{2}{c}{Expr} & \multicolumn{2}{c}{Accent} & \multicolumn{1}{p{0.1cm}}{}  \\
        \midrule

        \textsc{\absp}  &  & \multicolumn{2}{c}{\textbf{4.11 $\pm$ 0.08}} & \multicolumn{2}{c}{\textbf{4.00 $\pm$ 0.09}} & \multicolumn{2}{c}{{3.74 $\pm$ 0.09}} & \multicolumn{2}{c}{\textbf{4.00 $\pm$ 0.09}} & \multicolumn{2}{c}{\textbf{4.22 $\pm$ 0.08}} & \\

        \textsc{\ab} &  & \multicolumn{2}{c}{3.95 $\pm$ 0.08} & \multicolumn{2}{c}{3.97 $\pm$ 0.09} & \multicolumn{2}{c}{\textbf{3.88 $\pm$ 0.08}} & \multicolumn{2}{c}{{3.93 $\pm$ 0.09}} & \multicolumn{2}{c}{{4.17 $\pm$ 0.07}} &  \\
        \bottomrule
    \end{tabular}
    }
\end{table}

\begin{table}[ht]
    \centering
    \caption{Comparing unified \textsc{\ab} for Text-to-audio generation results on AudioCaps evaluation set. We find that \textsc{\ab} outperforms all baselines except the sound only \textsc{\absd}. Most notably it even outperforms TANGO-full-FT model on most metrics by significant margin.}
    \label{tab:ab_tta}
    \resizebox{\linewidth}{!}{
    \begin{tabular}{c|ccccc|cc}
        \toprule
                    & \multicolumn{5}{c|}{objective} & \multicolumn{2}{c}{subjective}\\
                    & FAD $\downarrow$ & FD $\downarrow$ & KLD $\downarrow$ & IS $\uparrow$ & CLAP $\uparrow$ & OVL $\uparrow$ & REL $\uparrow$\\
        \midrule
        ground truth & -	& -&	- &	\textbf{13.28} &	0.49  & 3.36\ci{0.18} & 3.86\ci{0.18} \\
        \midrule
        \multicolumn{8}{c}{Unified Models} \\
        \midrule
        VoiceLDM~\cite{lee2023voiceldm} & 10.28 &	49.48 &	2.95 &	4.79 &	0.37 & 2.07\ci{0.16} & 2.62\ci{0.22} \\
        UniAudio \cite{yang2023uniaudio} & 3.12 & - & 2.60 & - & - & - & -\\ 
        \textsc{\ab{}} (ours) & \bf{1.10} & \bf{10.14} & \bf{1.19} & \bf{11.90} & \bf{0.70}  & \textbf{3.19\ci{0.14}} & \textbf{3.94\ci{0.14}} \\
        \midrule
        \multicolumn{8}{c}{Sound-only models} \\
        \midrule        
        TANGO-full-FT~\cite{ghosal2023text} & 2.19 &	18.47 &	1.20 &	8.80 &	0.56 & 3.04\ci{0.13} & 3.78\ci{0.15} \\
        \textsc{\absd{}} (ours) & \textbf{0.77} &	\textbf{8.30} &	\textbf{1.15} &	\textbf{12.70} &	\textbf{0.71}  & \textbf{3.43\ci{0.15}} & \textbf{4.09\ci{0.15}} \\
        \bottomrule
    \end{tabular}
    }
\end{table}

\section{Inference Optimization with Bespoke Solver}

To generate samples from a flow-matching model, an ODE solver is used at inference time to approximate the integration. There are many solvers that one can choose from, such as adaptive step-size \texttt{dopri5} solver or fixed step-size \texttt{midpoint} solver. These solvers can be configured to operate at different speed-accuracy trade-off (accuracy in computing the integral). While flow-matching with OT path produces higher quality samples compared to diffusion models~\citep{flow-matching, le2023voicebox} for the same number of ODE steps and achieves better trade-off, very aggressive settings like \texttt{midpoint} with only 4 steps may still dramatically decrease the sample quality.

Inference efficiency is quantified by the number of function evaluation (NFE), which denotes the number of time an ODE solver evaluates the derivative.
To improve the inference speed at the extreme low NFE regime (i.e., 4), we adopt Bespoke Solvers \cite{shaul2023bespoke} to recover similar sample quality as the original model with a much lower NFE. 

Assume the initial noise sample $x(0) = x_0 \sim p(x_0)$. Bespoke solver learns extra parameters $\theta \in \mathbb{R}^{p}$ where $p$ is very small and minimize the \textit{global truncation error} (sum of \textit{local truncation error}) between approximate sample $x^{\theta}_n$ and ground truth data point $x(1)$ in the following formula: $\mathbb{E}_{x_0 \sim p(x_0)} \|x(1)-x_n^{\theta} \|$, where $x_n^\theta$ is the output of the solver step$^{\theta}$.

At a high level, Bespoke solvers aims to learn transformation for paths such that transformed can be more accurately estimated with the desired number of ODE steps. Bespoke Solver work by transforming the sample trajectory $x(t)$ using two components $t_r: [0,1] \rightarrow [0,1]$ as time reparameterization and invertible function $\varphi: [0,1] \times \mathbb{R}^d \rightarrow \mathbb{R}^{d}$, where those functions are parameterized by extra parameters $\theta$. 
Let the parametric solver be $\text{step}^{\theta}(t,x;u_t)$. 
First we transform input $(t, x)$ into $(r, \bar{x}) = (r_t, \varphi_{r_t}(x))$. 
Next, we perform a step in the transformed space as $(r_{next}, \bar{x}_{next}) = \text{step}(r, \bar{x}; \bar{u}_{r})$, using the chosen base solver (e.g., \texttt{midpoint}), where $\bar{u}_{r}$ is vector field on transformed trajectory. 
To transform back to original space, we compute $(t_{next}, x_{next}) = step^\theta(x,t;u_t) = (t_{r_{next}}, \varphi^{-1}_{r_{next}}(\bar{x}_{next}))$. 

To train the Bespoke solver, we generate the ground-truth path $x(t)$ at times $t_i$ where $i\in [N]$ using standard ODE solver, and we calculate the \textit{local truncation error} $d_i^{\theta} = \| x(t_i) - step_{x}^{\theta}(t_{i-1}, x(t_{i-1}); u)\|$ between ground truth and predicted sample from parameterized solver $\theta$, and finally we minimize the Bespoke loss $\mathcal{L}(\theta)=\mathbb{E}_{x_0 \sim p(x_0)} \sum_{i=1}^{n} d_{i}^{\theta}$. 

In this paper, we generate ground truth paths for training Bespoke Solvers for speech generation using \texttt{dopri5} ODE solver to estimate $N=200$ steps with guidance weight (GW) of 0.7. \cref{tab:inctx_bespoke} top half shows the evaluation result on zero-shot TTS with matched guidance weight (0.7) comparing two standard ODE solvers: \texttt{midpoint} and \texttt{dopri5} with the Bespoke Solver. 
As we can see, by using bespoke solver, we could reduce ODE steps down to 4 and still retain similar performance in term of style similarity and WER.

In addition, we also study if a Bespoke Solver trained for a specific guidance weight generalizes to a different guidance weight, and present comparison between the default \texttt{midpoint} solver with the bespoke solver using GW=0.0. Results suggest that it can generalize to different guidance setups.

\begin{table}[ht]
    \centering
    \caption{Comparison between the standard ODE solver using midpoint, \texttt{dopri5} and parameterized Bespoke solver in term of NFE, speaker similarity and WER.}
    \label{tab:inctx_bespoke}
    \resizebox{\linewidth}{!}{
    \begin{tabular}{ccc|ccccc|ccccc}
        \toprule
        Solver & NFE & GW & \multicolumn{5}{c|}{Sim-o $\uparrow$} & \multicolumn{5}{c}{Word error rate (\%) $\downarrow$} \\
        & & & LS & CV & SWBD & Expr & Accent & LS & CV & SWBD & Expr & Accent \\
        \midrule
        \texttt{dopri5}       & $\sim$280 & \multirow{3}{*}{0.7} & 0.733 & 0.607 & 0.605 & 0.602 & 0.657 & 3.0 & 3.6 & 9.5 & 2.8 & 0.9 \\
        \texttt{midpoint}, 16 steps & 32        & & 0.734 & 0.607 & 0.608 & 0.603 & 0.659 & 3.2 & 3.7 & 9.1 & 3.2 & 0.9 \\
        Bespoke, 4 steps  & 8         & & 0.735 & 0.607 & 0.606 &	0.606 & 0.658 & 3.0 & 3.5 & 8.3	& 3.0 & 0.7 \\
        \midrule
        \texttt{midpoint}, 16 steps & 32        & \multirow{2}{*}{0.0} & 0.671 & 0.546 & 0.578	& 0.541	& 0.601 & 3.6 &	5.1	& 12.1 & 3.1 & 1.3 \\
        Bespoke, 4 steps  & 8         &                      & 0.672 & 0.548 & 0.576	& 0.544	& 0.604 & 3.6 & 5.1 & 12.1 & 3.0 & 1.3 \\
        \bottomrule
    \end{tabular}
    }
\end{table}
\section{Responsible AI}
In order to build a system responsibly, we conduct evaluations to gauge the fairness aspect and studies methods to defend misuse. In this section, we first analyze if our model produces similar performance on different groups like genders and accents. Second, we also perform watermarking experiments to evaluate if a recently proposed watermarking system generalizes to our models such that watermarked samples from our models can be reliably detected.

\subsection{Fairness across groups}
We train our model on large quantities of data from various sources. We believe through scaling training data, our model can perform well across many different groups. We assess this aspects by evaluating model performance by genders and by accents. In particular, we consider gender bias or accent bias are observed if there is a groups that has significantly worse performance in term of content correctness (measured by WER) and style similarity (measured by cosine similarity between style embeddings) compared to those of the entire population.

To conduct our experiment, we consider the zero-shot TTS task conditioned on a context prompt. We use a dataset with country and gender labels for this experiment. For the TTS transcript, we sample 20 transcripts from the test set. For the TTS prompts, we evaluate on accents of which there are at least 5 unique speakers in the dataset, which leave us to 64 accents. Then, we sample 20 random utterances (10 for male, 10 for female) from each accent groups. In total, we have 400 (20 transcripts $\times$ 20 prompts) for each accent groups and 12800 (20 transcripts $\times$ 10 prompts $\times$ 64 accents) for each gender groups.

\begin{figure}[h]
\centering
\begin{subfigure}[t]{0.4\textwidth}
\centering
\includegraphics[width=\linewidth]{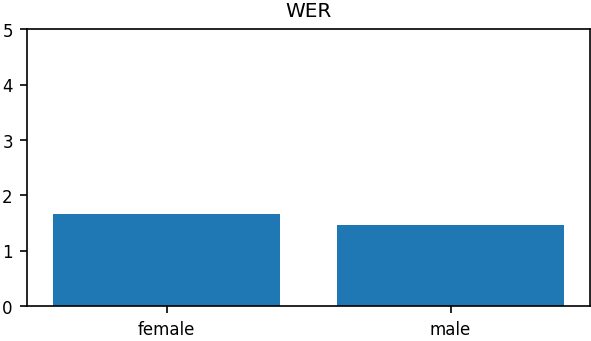}
\caption{WER across gender group.} \label{fig:fairness_gender_wer} 
\end{subfigure}\hspace{1em}
\begin{subfigure}[t]{0.4\textwidth}
\centering
\includegraphics[width=\linewidth]{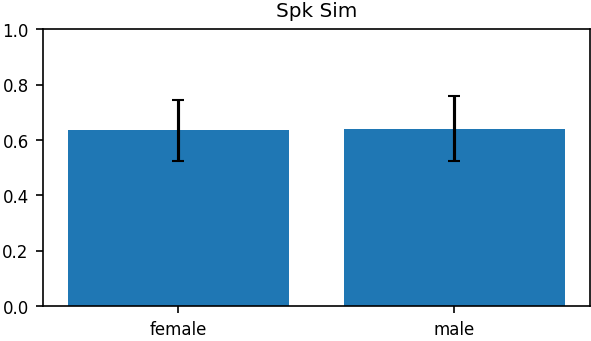}
\caption{Speaker similarity across gender group (mean $\pm$ 1 stddev).} \label{fig:fairness_gender_spksim}
\end{subfigure}
\end{figure}

\begin{figure}[h]
\begin{subfigure}{\textwidth}
\centering
\includegraphics[width=\linewidth]{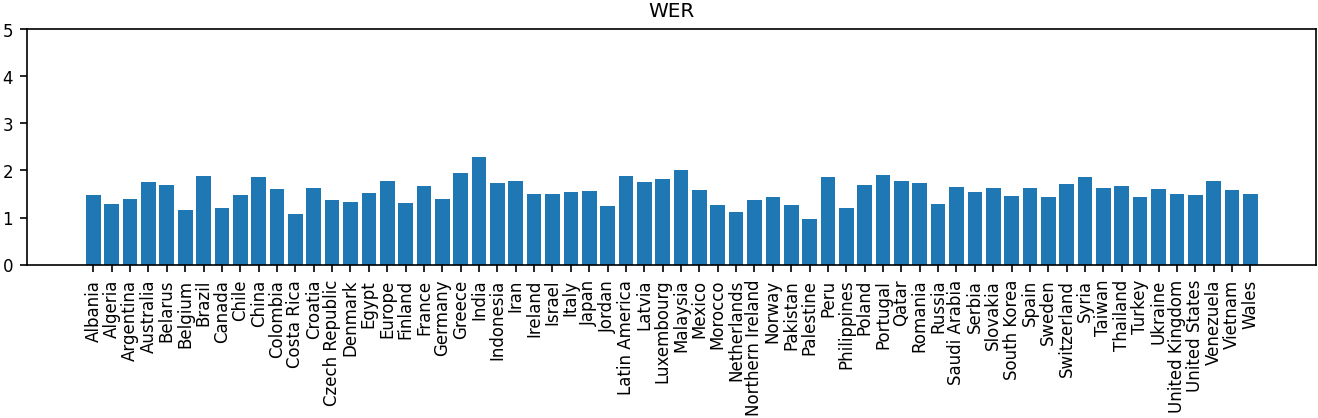}
\caption{WER across accent group.} \label{fig:fairness_accent_wer}
\end{subfigure}
\begin{subfigure}{\textwidth}
\centering
\includegraphics[width=\linewidth]{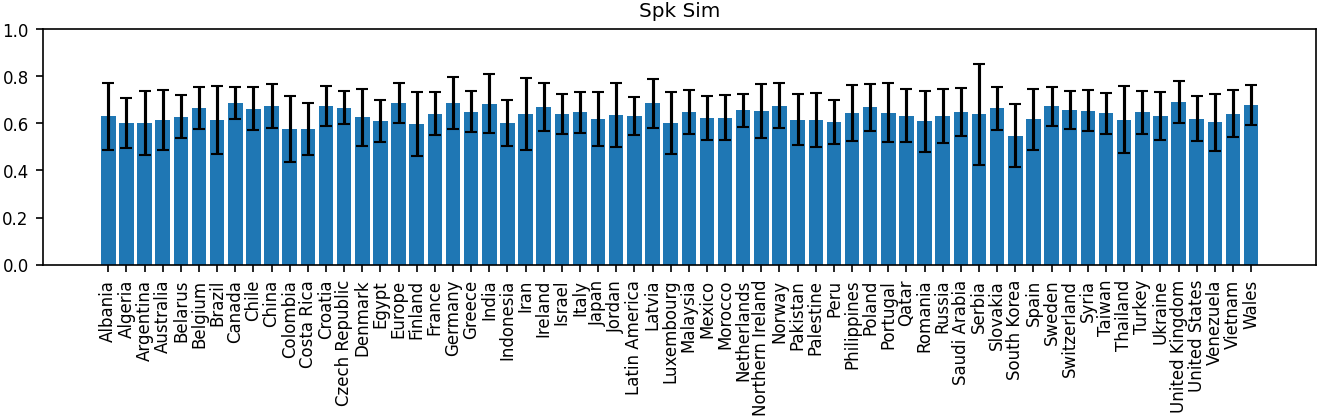}
\caption{Speaker similarity across accent group (mean $\pm$ 1 stddev).} \label{fig:fairness_accent_spksim}
\end{subfigure}
\end{figure}

\cref{fig:fairness_gender_wer} shows average WER and \cref{fig:fairness_gender_spksim} shows average speaker similarity across different gender group. We observed that the number are very similar and the speaker similary mean fall between $\pm$ 1 standard deviation. \cref{fig:fairness_accent_wer} shows average WER and \cref{fig:fairness_accent_spksim} shows average speaker similarity across different accent group. Similar with the gender groups, WER over all accents remain similar and each group speaker similarity falls within $\pm$ 1 standard deviation. Across gender and accent, WER remains very low around 1.5\% which means 1 mistake for every 66 words in the transcript. We come to the conclusion that our model has no significant performance difference given different group of gender and accents.

\subsection{Watermarking for Generated Audio Detection}
Recent advancement on quality and fidelity in audio generative model has empower novel applications and use case on the model. However, at the same time, there are many people has their raising concerns about the risks of misused. Therefore, the ability to recognize which audio is generated or real is crucial to prevent the misused of the technology and enable certain platform to comply with their policy \cite{Fernandez_2023_ICCV}.

In this section, we use Seamless Watermark~\citep{seamlessv2} to see we can reliably put and detect an imperceptible watermark on top of our model generated audio. The watermarking model has similar building block as Encodec \cite{defossez2022highfi}. The training objectives are based on weighted combination of two losses: 1) perceptual loss to ensure the watermark is imperceptible (Si-SNR and L1 loss), 2) localization loss based on binary cross entropy to ensure accurate localized detection on watermark in frame level. 

Here, we use the output generated from most scenarios such as zero-shot TTS, description-based TTS, voice+description-based TTS, and sound generation and apply various data augmentation on top of them. We measure the performance of watermark detection by their false positive rate (FPR) and false negative rate (FNR).

\begin{table}[h]
\centering
\begin{tabular}{c|c|c}
\hline
\toprule
\textbf{Augmentation}    & \textbf{FPR}   & \textbf{FNR}   \\
\midrule
No augmentation & 0.001 & 0     \\
Bandpass filter & 0.001 & 0     \\
Boost audio     & 0.001 & 0     \\
Duck audio      & 0.001 & 0     \\
Echo            & 0.001 & 0.001 \\
Highpass filter & 0.001 & 0     \\
Lowpass filter  & 0.001 & 0     \\
Pink noise      & 0.001 & 0     \\
Random noise    & 0     & 0     \\
Speed slower    & 0     & 0.003 \\
Smoothing       & 0     & 0.001 \\
Up-down resampling & 0.001 & 0  \\
\bottomrule
\end{tabular}
\caption{List of audio augmentation technique applied on top of watermarked audio with their detection performance respectively averaged on all scenarios.} \label{tab:watermark}
\end{table}

\cref{tab:watermark} shows the average FPR and FNR over all tasks for each data augmentations. We observed very low FPR and FNR, close to 0\%, which means the watermark works very robustly against various type of generated audio and speech and data augmentations. Simultaneously, the watermarked audio also have very low scale-invariant signal-to-noise ratio (SI-SNR) -20.6db, which means the watermarks residual is in-perceivable from human perspective.
\section{Discussion}

\subsection{Limitations}

\boldparagraph{Fine-grained Control}
With the recent advances in generative models, the performance in terms of controllability is mainly determined by the domain coverage and the quantity of the training data. We have demonstrated that for in-context TTS (example-based control), style similarity can be significantly improved by scaling the data used for self-supervised pre-training, which learns to infill audio given the audio context.

In contrast, description-based control requires a higher level of supervision, using paired audio and description to align concepts described in text with variations observed in audio. Hence, it is harder to generalize description-based control due to the scarcity of labeled data covering various concepts and concepts of different granularity. 

To give a concrete examples, our training data may contain both instances of chihuahua barking and those of labrador barking; however, all those instances are likely captioned as ``a dog barking.'' Hence, when prompted with ``a chihuahua barking,'' the best the model can do is generating a dog barking audio clip if the text embedding of ``chihuahua'' and ``dog'' are close to each other, but it would not be able to generate the correct chihuahua barking sound if such supervision was not provided during training. The same idea also applies to speech attributes such as accents, where regional accents cannot be accurately generated if the training dataset does not include those paired examples.

\boldparagraph{Data creation}
Given the coverage and the quantity of paired data is the key to improve description-based control, it is natural to consider strategies to create such data. However, it is in fact very challenging to create fine-grained descriptions given audio. 
While it is easy for annotators to differentiate cat meowing and dog barking, labeling which dog species solely based on audio is difficult task for most of the people. Similar challenges exist as well regarding labeling speech attributes such as accents. Moreover, annotators can often disagree on attributes such as emotion, perceived age and quality of audio. 
Hence, it is difficult to create large scale fine-grained description datasets for audio. The lack of large such datasets also leads to difficulty in developing attribute taggers and captioning models that can automate description creation and be used for evaluation.

\subsection{Broader Impact}
This work greatly advances controllability for speech generation and improves coverage of styles. The ability to generate speech with desired vocal and acoustic styles using natural language descriptions unlocks a wealth of applications. For example, it can be used to create new voices for characters in immersive audiobooks, Ads, and movie scripts, where creators have in mind what style of voice the characters should have. Compared to exampled-based control (in-context TTS), description-based control can create novel voice of the desired without having to clone from an existing individual and saves the time creators spends on searching the reference voice.

The ability to generate speech in diverse acoustic conditions is especially crucial for applications such as film making and immersive audiobook creation, where characters may be presented in different environment such as caves and it is essential to create audio reflecting the acoustic properties of those scenes.
The ability to preserve the voice while changing emotion and acoustic scenes is also crucial for generating long form audio content such as stories. Overall, \ab{} makes it much easier for creators to generate content with higher quality compared to prior models.

While \ab{} can help spark everyone's creativity and bring many positive social impacts, similar to other powerful generative models, it also carries risks of being misused and causing unintended harm. In particular, speech synthesis may be used for spreading misinformation and impersonation. We presented studies on watermarking to effectively mitigate this risk in a robust fashion. On other hand, we also demonstrated that model perform similarly well across variations demographic groups, ensuring bias is reduced through data scaling.
\section{Conclusion}
This paper presents \textsc{\ab{}}, a unified model for audio generation with unprecedented versatility, controllability, and quality.
\textsc{\ab{}} is capable of generating both speech and sound from text description, audio example, or a combination of vocal style reference and description. In particular, for speech generation, \textsc{\ab{}} is able to control very fine-grained vocal styles such as accent, emotion, timbre and create speech simulating more diverse environment compared to previous models. Asides from showing novel capabilities, \textsc{\ab{}} outperforms all prior in-context speech generation and sound generation models on well-studied benchmarks evaluating existing capabilities.

More importantly, we believe this work pioneers in building universal audio generative models with unified controls and sheds light for future research on audio generative modeling. In essence, we demonstrate that with large quantities of data, it is possible to build a unified model that outperforms modality specific ones. This points toward a path similar to the evolution of language generation models, where a large scale model trained with a simple objective on large quantities of data eventually surpasses task or language specific models with significantly better generalization ability and emerging capabilities. 
\section*{Acknowledgement}

The authors would like to 
thank Ricky Chen, Hady Elsahar, Ilia Kulikov, Hirofumi Inaguma, Jing Xu, and Yossi Adi, Alexander H. Liu, Chung-Ming Chien, Qing He, Thilo Koehler, Fuchun Peng, Xiaohui Zhang, Vimal Manohar, Po-Wei Chou, Kaustubh Kalgaonkar, Anurag Kumar, Yangyang Shi, Zhaoheng Ni, Gael Le Lan, and Varun Nagaraja for their helpful discussion on research,
thank Juan Pino, Ian Stewart, Alexander Miller, and Joelle Pineau for organizational support,
thank Adina Williams, Christophe Ropers, Chloe Bakalar, Imanol Arrieta Ibarra, and Esteban Arcaute for discussion on responsible AI,
thank Wei Zhu, Yichen Wang, Jiun-Ren Lin, Chao Zhou, Peter Weng, Stephen Fink, Ivan Evtimov, David Renardy, Sonia Kim for responsible AI and safety implementation,
thank Neil Seejoor, Somya Jain, Chandan Avdhut, Chris Henry, and KC Braunschweig for support on infrastructure,
thank Carolyn Krol, Ernest Hammond, Mo Metanat, David Soofian, Ndidi Elue, Mallika Malhotra, Kelechi Ebi Kamanu, Maeve Ryan, Harrison Rudolph, Jennifer Okafor for their support on research and grant review,
thank Ana Paula Kirschner Mofarrej, Lydia Baillergeau, Steph Miles, Raghu Nayani, Michelle Restrepo, Tamara Piksa, Chris Wiltz, Orialis Valentin, Aiman Farooq, Gopika Jhala and Ashley Gabriel on cross-functional support

\clearpage
\section*{Contribution}
\textbf{Apoorv Vyas} proposed and implemented LLM caption, audio augmentation and annotation quality control pipelines, and implemented voice prompting

\textbf{Bowen Shi} led Audiobox-Sound experiments, implemented and conducted experiments for Joint-CLAP, proposed two-stage fine-tuning and led studies on evaluation

\textbf{Matthew Le} implemented and conducted experiments for Audiobox-SSL, Audiobox-Speech, and Bespoke Solver, led model integration to demo 

\textbf{Andros Tjandra} implemented speech attribute labelers and responsible AI studies

\textbf{Yi-Chiao Wu} created Audiobox baseline results and implemented audio infilling for baselines

\textbf{Liang Tan} explore speech representation and conducted preliminary experiments on forced aligners

\textbf{Bowen Shi and Wei-Ning Hsu} prepared sound data and implemented, proposed Joint-CLAP and conducted experiments for Audiobox-Sound

\textbf{Andros Tjandra and Apoorv Vyas} implemented Audiobox

\textbf{Andros Tjandra and Matthew Le} conducted experiments for duration models

\textbf{Apoorv Vyas, Andros Tjandra and Bowen Shi} iterated on LLM prompting for text-to-speech and sound training

\textbf{Apoorv Vyas, Andros Tjandra, Matthew Le, Bowen Shi, Liang Tan and Wei-Ning Hsu} prepared speech data

\textbf{Apoorv Vyas, Andros Tjandra, Matthew Le and Bowen Shi} conducted Audiobox experiments

\textbf{Wei-Ning Hsu, Bowen Shi, Apoorv Vyas, Andros Tjandra, Matthew Le} wrote the paper 

\textbf{Baishan Guo, Apoorv Vyas and Andros Tjandra} implemented the human annotation pipeline

\textbf{Baishan Guo} ran human annotation and subjective evaluation, and analyzed annotation and evaluation results

\textbf{Bapi Akula} explored audio pre-processing and transformation, assisted in developing data pipeline

\textbf{Carleigh Wood} coordinated and facilitated data annotation

\textbf{Jiemin Zhang} led the demo development, designed and implemented demo infra, model integration, early demo mitigation and capacity testing

\textbf{Xinyue Zhang} designed and implemented demo backend, data logging, mitigation verification and toxic content filtering. 

\textbf{Robbie Adkins} designed and implemented demo frontend and supported backend implementation. 

\textbf{Akinniyi Akinyemi} conducted demo deployment, demo and mitigation infra set up.

\textbf{Joshua Lane} implemented early UI structure.

\textbf{William Ngan} designed the demo experience and implemented front-end demo interfaces.

\textbf{Brian Ellis} prototyped demo concepts and created audio for demos

\textbf{Alice Rakotoarison, Chris Summers} conducted demo user experience research

\textbf{Yael Yungster} provided design management support

\textbf{Jeff Wang} provided product management support for the team, contributed to overall research vision, strategy, project milestones and execution.

\textbf{Ivan Cruz} provided technical program management support, coordinated responsible AI study, red teaming, and cross-functional support

\textbf{Rashel Moritz} provided program management support, contributed to early project planning, mitigation planning, review, and cross-functional support

\textbf{Mary Williamson} provided management support for the team and co-led the project, contributed to research vision, and oversaw demo

\textbf{Wei-Ning Hsu} designed and led the project, advised Apoorv, Bowen, Matthew, Andros, Yi-Chiao, and Liang on the research, and coordinated research, demo, and data streams.

\bibliography{main}
\bibliographystyle{abbrvnat}


\clearpage
\appendix

\setcounter{table}{0}
\renewcommand{\thetable}{\Alph{section}\arabic{table}}
\setcounter{figure}{0}
\renewcommand{\thefigure}{\Alph{section}\arabic{figure}}

\section{Speech Attributes for Speech Caption Creation}\label{sec:app_sp_attr}

As described in \cref{sec:ab_spcap}, we extract attributes for creating speech captions. We obtain speech attributes from the associated metadata or by pseudo-labeling for a subset of attributes which can be labeled more reliably. Details for each attribute are listed below

\begin{itemize}
    \item Age: We first bin the age into 4 different categories namely less than twenty (<20),  young adults (20-35),  middle age (40-60), and elders (>60). We then fine-tune our dataset from pre-trained WavLM-base checkpoint with 3200 hours speeech and age metadata from our training set (consisted of conversational and reading speech with various quality).
    \item Gender:  We fine-tune on top of WavLM-base checkpoint with 4300 hours speech and gender metadata from our training set  (consisted of conversational and reading speech with various quality).
    \item Audio Quality: We use TorchAudio-Squim \cite{Kumar2023TorchAudioSquim} library and extract Perceptual Evaluation of Speech Quality (PESQ) \cite{Rix2001PESQ} score. We then bin the score into three categories: Low quality ( 0-2.39 ), Normal quality ( 2.39-3.8 ) and Studio Quality ( >3.8 ).
    \item Pitch: We use PyWorld vocoder \footnote{https://github.com/JeremyCCHsu/Python-Wrapper-for-World-Vocoder} to extract fundamental frequency (f0) and then calculate the geometric mean across all voiced region. We use gender dependent threshold for binning the pitch into three different categories: low, normal, high. For gender masculine, we set low pitch (0-40 percentile), normal pitch (40-90 percentile) and high pitch (>90 percentile). For gender feminine, we set low pitch (percentile 0-10), normal pitch (10-60 percentile) and high pitch (>60 percentile). The logic behind asymmetric threshold is because in general people will perceive most of masculine voice have lower pitch and most of feminine voice have higher pitch.
    \item Speaking rate: Given the transcript and audio, we first apply VAD to remove the silence segments. We then calculate character per seconds (CPS) and bin them into 3 categories: slow (<9.2 CPS), high (>20.8 CPS) and normal (9.2 <= x <= 20.8 CPS).
    \item Accent: We use the accent from the metadata whenever available in the metadata, otherwise leave it blank.
    \item Emotion: We use the emotion labels whenever available in the metadata, otherwise we leave it as blank.
    \item Environment: We use the environment tags such as inside a room, outside whenever available from the datasets.
\end{itemize}

\section{Automatic Captions: Quality}\label{sec:app_caption_quality}

To ensure we get high quality descriptions, we deploy a two-stage approach to filter the annotator candidate. First, we keep only annotators that successfully labeled pre-selected gold samples with high accuracy (> 73\%). Later, we score their submitted captions using LLM and keep the annotator if their averaged score is above certain threshold. More specifically, for a speech segment, we first use a LLM to generate a caption based on annotated audio attributes. We then run a second stage where we ask the  another LLM to compare the LLM-generated caption with human-written caption and rate the quality of human-written captions from 1 to 5. We prompt this LLM to give low score to human-written captions where no interesting audio events were added in additional to the annotated audio attributes or some important audio attributes are missing. Annotators with an averaged caption score less than 3 were removed. This resulted in high quality and detailed captions that complement our pseudo-labeled captions above. Here are some captions example curated by our human annotator:

\section{Unified Audiobox Task Description}\label{sec:app_task_inputs}

Below we describe different tasks that unified \textsc{\ab{}} model can solve along with the inputs required.
\begin{itemize}
    \item Zero-shot TTS (in-context TTS): model takes as input a transcript and an audio example and generates speech that resembles the example's audio style (described in \cref{sec:ab_sp_task}). Inputs: (context, transcript).
    \item Description-TTS/TTA: model takes as input a transcript/pseudo-transcript and a text description and generates speech / audio matching the description. Inputs: (description, transcript/pseudo-transcript)
    \item Voice Restylization: model receives a transcript, a voice prompt and a description. The generated output needs to match speaker's vocal style and the description. Note that the description could contain attributes different from the voice prompt. For the voice prompt could have been recorded in a small room with neutral emotion and the description could specify happy emotion in a church with bells ringing. Inputs: (voice, description, transcript)
    \item Sampling: The model receives a transcript as input and samples diverse voices. Inputs: (transcript)
    \item Speech Infilling/Editing: model takes as input an masked speech with accompanying transcript and an optional description and infills the masked portion. Inputs: (context, transcript, optional description)
    \item Audio Infilling/Editing: model takes as input an masked audio with pseudo transcript and description to infill the masked portion with matching description. Inputs: (context, pseudo transcript, description)
\end{itemize}

\section{Subjective Evaluation Interface}\label{sec:app-mos-eval}

\begin{figure}
  \centering

  \begin{minipage}[b]{0.75\textwidth}
    \includegraphics[width=\textwidth]{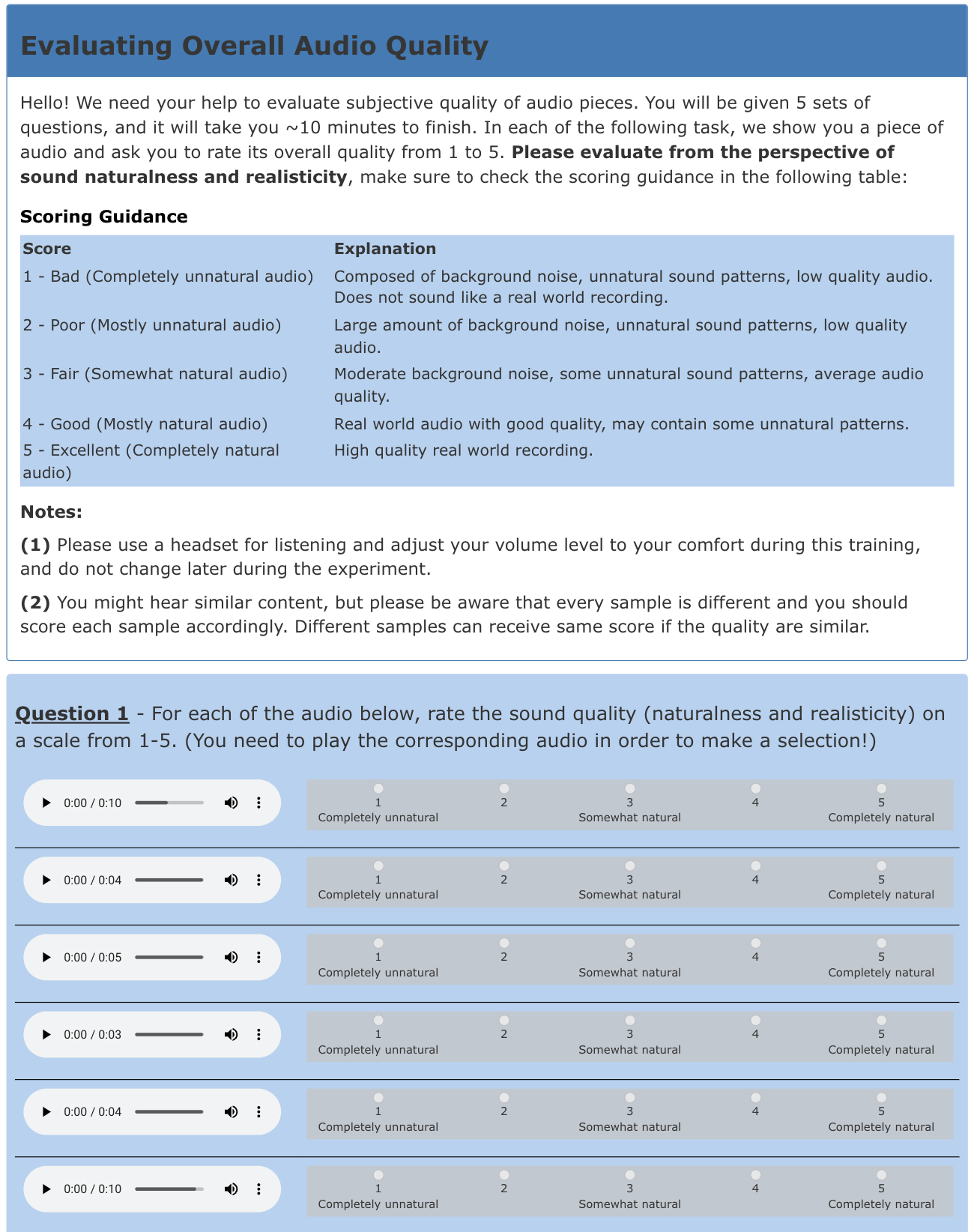}
    \caption{OVL evaluation for sound}
    \label{fig:sound-ovl-interface}
  \end{minipage}
\end{figure}

\begin{figure}
  \centering

  \begin{minipage}[b]{0.75\textwidth}
    \includegraphics[width=\textwidth]{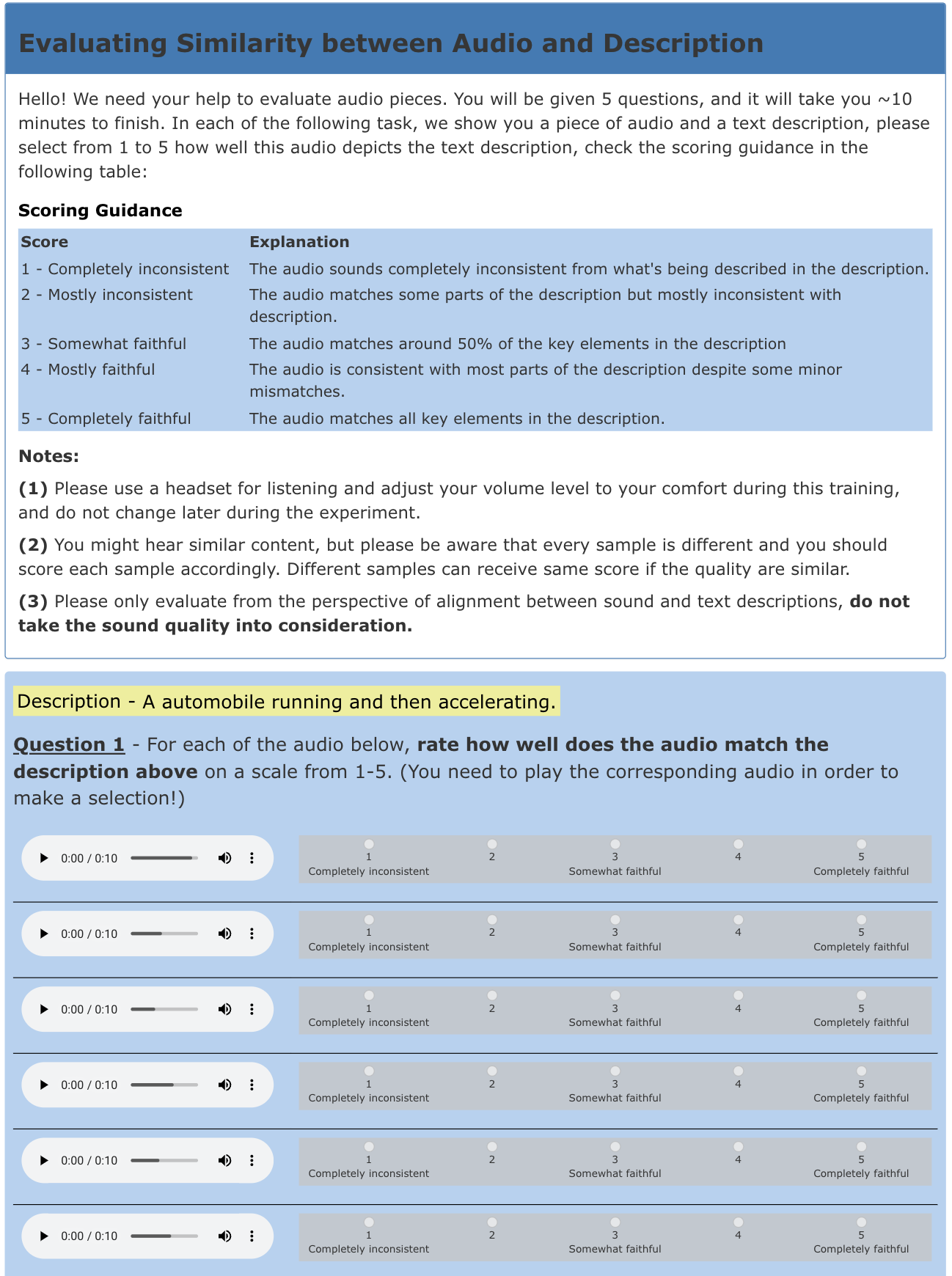}
    \caption{REL evaluation for sound}
    \label{fig:sound-rel-interface}
  \end{minipage}
\end{figure}

\begin{figure}
  \centering

  \begin{minipage}[b]{0.75\textwidth}
    \includegraphics[width=\textwidth]{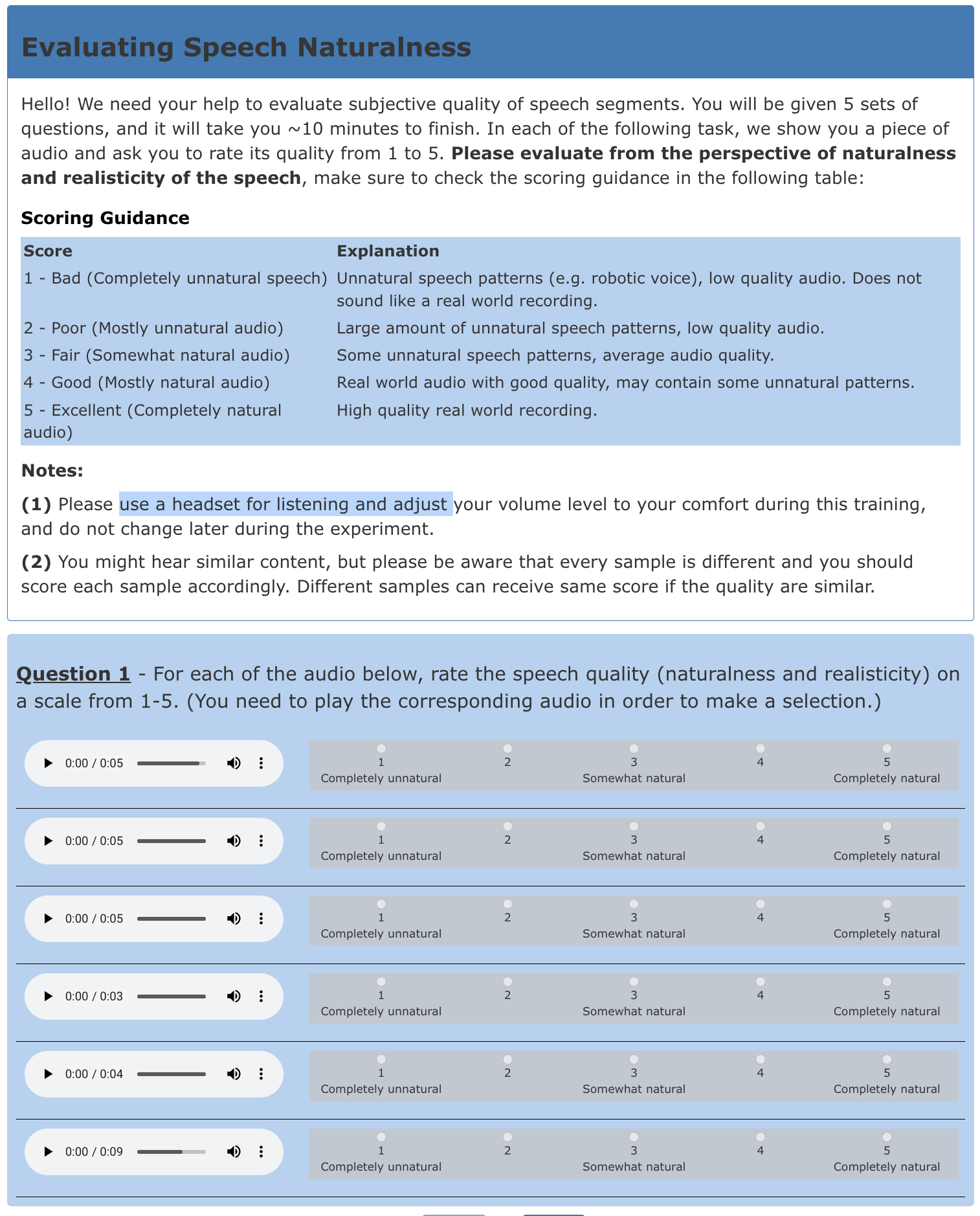}
    \caption{Quality MOS evaluation for speech}
    \label{fig:speech-ovl-interface}
  \end{minipage}
\end{figure}

\begin{figure}
  \centering

  \begin{minipage}[b]{0.75\textwidth}
    \includegraphics[width=\textwidth]{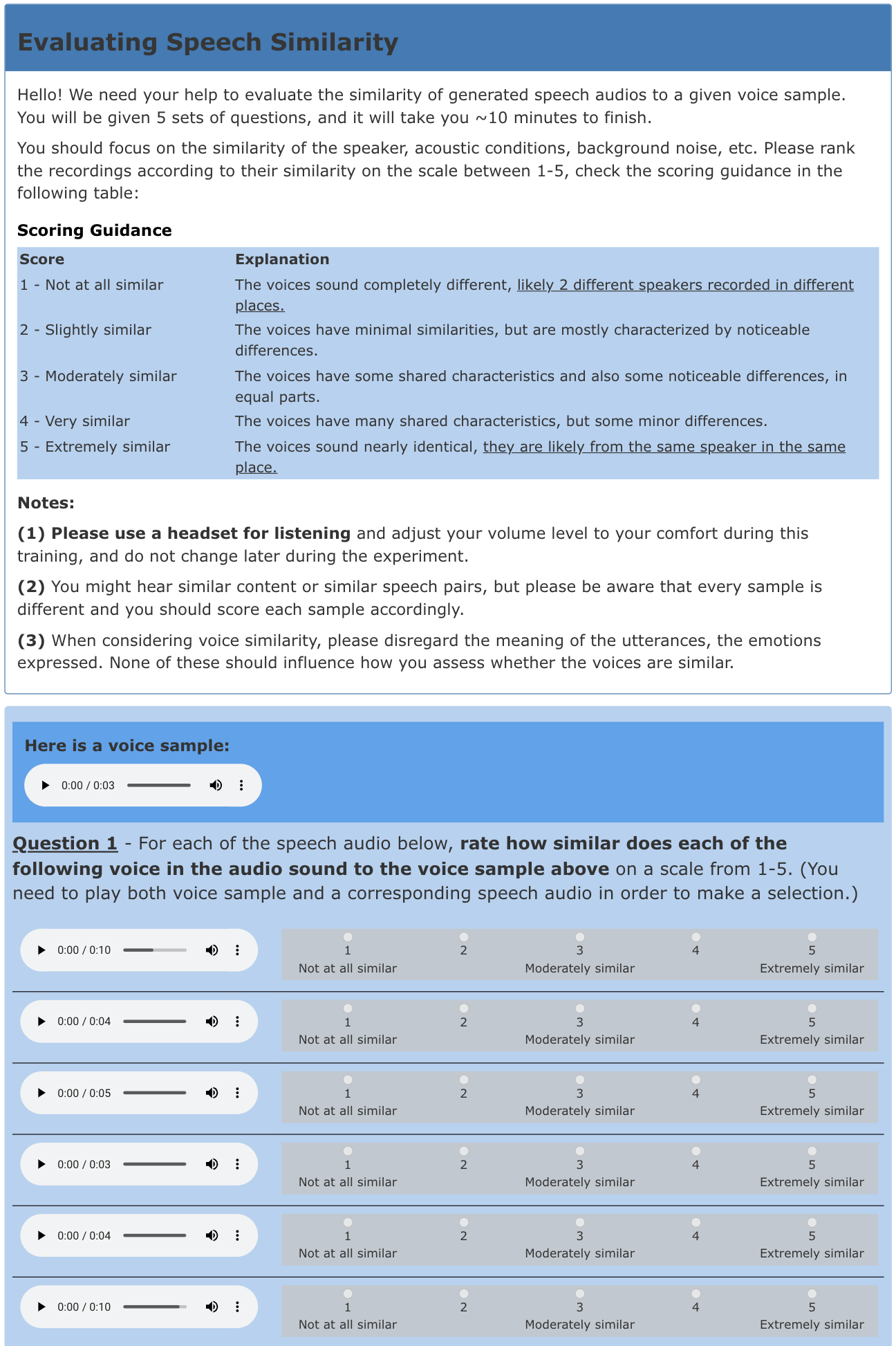}
    \caption{Similarity MOS evaluation for speech}
    \label{fig:speech-smos-interface}
  \end{minipage}
\end{figure}

\begin{figure}
  \centering

  \begin{minipage}[b]{0.75\textwidth}
    \includegraphics[width=\textwidth]{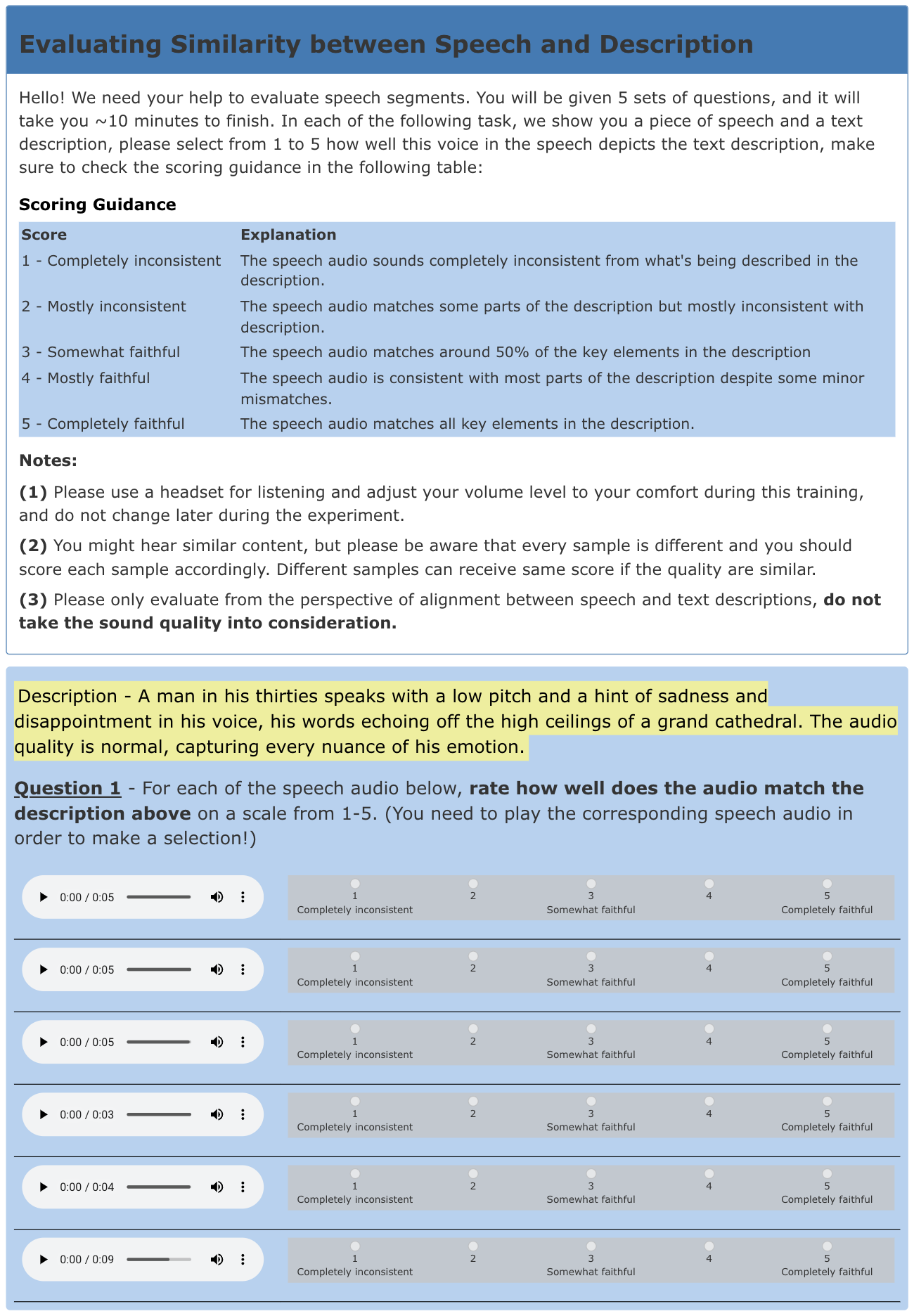}
    \caption{REL evaluation for speech}
    \label{fig:speech-rel-interface}
  \end{minipage}
\end{figure}

We show human annotation interfaces for sound in \cref{fig:sound-ovl-interface,fig:sound-rel-interface}, for speech in \cref{fig:speech-ovl-interface,fig:speech-rel-interface,fig:speech-smos-interface}. 

\end{document}